\documentclass[aps,twocolumn]{revtex4}
\usepackage{epsfig}

\newcommand{\be}{\begin{equation}}
\newcommand{\ee}{\end{equation}}
\newcommand{\ba}{\begin{eqnarray}}
\newcommand{\ea}{\end{eqnarray}}

\usepackage{epsfig}
\begin{document}

\title{What is the order of 2D polymer escape transition?} 
\author{Hsiao-Ping Hsu and Kurt Binder}
\affiliation{Institut f\"ur Physik, Johannes Gutenberg-Universit\"at Mainz\\
D-55099 Mainz, Staudinger Weg 7, Germany}
\author{Leonid I. Klushin}
\affiliation{American University of Beirut, Department of Physics, 
Beirut, Lebanon}
\author{Alexander M. Skvortsov}
\affiliation{Chemical-Pharmaceutical Academy, Prof. Popova 14, 197022 
St. Petersburg, Russia.}

\date{\today}

\begin{abstract}

An end-grafted flexible polymer chain in 3d space between two pistons 
undergoes an abrupt transition from a confined coil to a flower-like 
conformation when the number of monomers in the chain, $N$, reaches 
a critical value. In $2d$ geometry, excluded volume interactions between 
monomers of a chain confined inside a strip of finite length $2L$ transform 
the coil conformation into a linear string of blobs. However, the blob 
picture raises questions on the nature of this escape transition. 
To check the theoretical predictions based on the blob picture 
we study $2d$ single polymer chains with excluded volume interactions 
and with one end grafted in the middle of a strip of length $2L$ and 
width $H$ by simulating self-avoiding walks on a square lattice with 
the pruned-enriched-Rosenbluth method (PERM). We estimate the free energy, 
the end-to-end distance, the number of imprisoned monomers, 
the order parameter, and its distribution. It is shown that in the 
thermodynamic limit of large $N$ and $L$ but finite $L/N$, there is a 
small but finite jump in several average characteristics, including 
the order parameter. We also present a theoretical description based 
on the Landau free energy approach, which is in good agreement with 
the simulation results. Both simulation results and the analytical 
theory indicate that the $2d$ escape transition is a weak first-order 
phase transition.
\end{abstract}

\maketitle

\section{Introduction}

A phenomenon that was called escape transition occurs upon
progressive squeezing an end-grafted polymer chain between two pistons
and has attracted great interest [1-16]. At weak
deformation the chain is compressed uniformly into a relatively thick
pan-cake conformation. Beyond certain critical compression, the chain
configuration changes abruptly. One part of the
chain forms a stem stretching from the grafting point to the piston edge, 
while the rest of the segments form a coiled crown outside the piston,
thus escaping from the region underneath the piston.
An abrupt change from one state to another implies a first order transition.
Various aspects of this problem were investigated:
The escape transition of compressed polymer mushrooms in 3d space was
investigated thoroughly by scaling theory~\cite{Subra},
numerical calculations~\cite{Ennis99,Sevick99,Steels},
and computer modeling under good solvent~\cite{Milchev},
and theta solvent~\cite{Jimenez} conditions. 
The escape transition of star polymers
was discussed in~\cite{Sevick00}, and the escape transition of 
di-block-copolymers was considered in~\cite{Ennis01}. 
The influence of the curvature of the pistons was investigated
in~\cite{Williams95, Guffond97}, and the effect of adsorption between
the polymer chain and the surface of the piston was considered 
in~\cite{Leermakers02}.
A comparison between Monte Carlo simulations and experimental results
by atomic force-electrochemical microscopy
was recently presented in~\cite{Abbou}.
A rigorous analytical theory for the equilibrium and kinetic
aspects of the escape transition for a Gaussian chain was constructed 
in~\cite{Skvortsov02, Klushin}. Metastability effects, negative compressibility
and the nonequivalence of the escape transition
in two conjugate ensembles were analyzed for the same model 
in~\cite{Leermakers04, Skvortsov06}.

    The reason for studying the escape transition is that it
gives the possibility to understand the phenomenon of a very 
unconventional phase transition.
The concept of a phase transition requires taking a thermodynamic limit.
For standard low-molecular weight systems as well as for macromolecular systems
in condensed bulk matter finite size effects are usually negligible. In
contrast to that, phase transitions at the level of a single macromolecule
e.g., the coil-globule transition~\cite{Grosberg}, or polymer adsorption at
an interface~\cite{Eisenriegler, Fleer} - do not have any analogies
in the physics of low molecular mass systems. A single
macromolecule always consists of a finite number of monomers $N$:
computer modeling rarely deals
with $N$ larger than $10^{4}$ so that finite-size effects in the
single-molecule phase transitions are the
rule rather than the exception. 
The situation is much more complicated in the case of the escape transition.
It was shown~\cite{Subra} that the escape transition
point (critical compression) 
depends on the relation between the chain length $Na$ ($a$ is the distance
between neighboring monomers) and the piston radius $L$.
Therefore, to analyze the escape transition in the thermodynamic limit,
it is necessary to take both $Na\rightarrow\infty$ and $L\rightarrow \infty$
but $Na/L=const$.
                                                                               
The physics of phase transitions is generally known to be strongly
affected by the spatial dimensionality. For phase transitions at the level
of a single macromolecule the spatial
dimensionality is important because excluded volume effects are especially
large for polymeric
chains in two dimensions~\cite{deGennes}. An ideal chain without
excluded volume interactions retains
a Gaussian coil conformation with the lateral size $\sim N^{1/2}$ even
when it is confined in a 2d strip.
There is a very pronounced difference between this state and the
partially escaped state with a strongly stretched stem of 
size $L \sim Na$.
On the contrary, a 2d polymer chain with excluded volume
interactions confined in a strip is already strongly elongated with
the size $\sim Na$ . It is not clear
whether the difference between the confined and the escaped states is
large enough to result in a phase transition.

\begin{figure}
\begin{center}
\epsfig{file=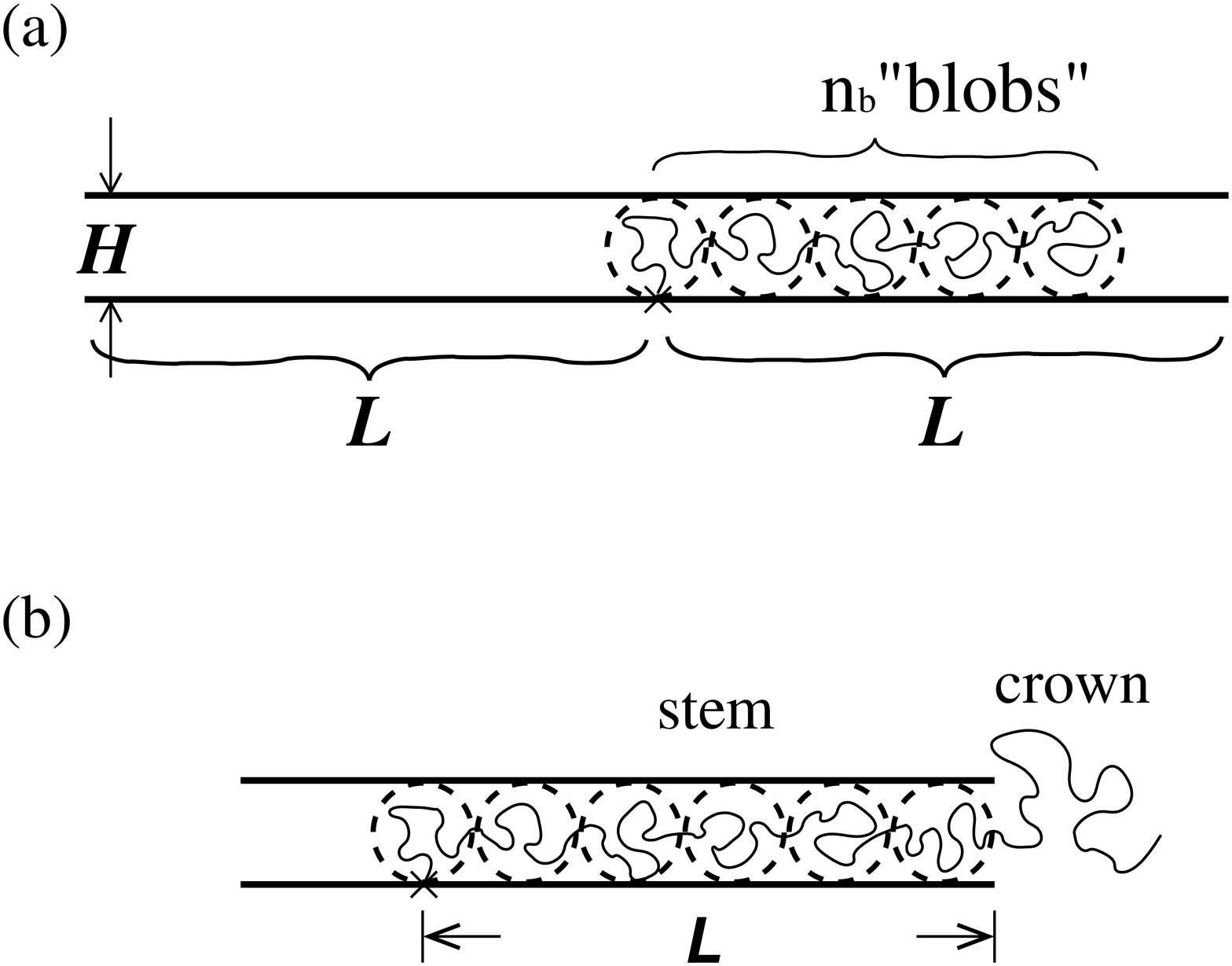, width=8.0cm, angle=0}
\caption{Schematic drawings of a flexible polymer chain of length $N$ grafted
in the middle of the strip of length $2L$ and width $H$, in a blob picture:
(a) As the chain
is imprisoned inside the strip, it forms a sequence of $n_b$ blobs.
(b) As the chain length is larger than the maximum chain length $N^*$
of a chain in an imprisoned state, the chain partially escapes from the
strip and forms an escaped state. A escaped state consists of
a ``stem" containing $N^*$ monomers and a ``crown" containing
$N-N^*$ monomers.}
\label{fig-blobc}
\end{center}
\end{figure}

To analyze the excluded volume effects of the escape transition,
we study a flexible polymer chain containing $N$ links of
length $a$ ($N$ monomers) grafted in the middle of a strip of length $2L$ and width $H$,
under good solvent conditions. Schematic drawings of a polymer chain in an
imprisoned state and in an escaped state are shown in Fig.~\ref{fig-blobc}.
First, we summarize the results of the escape transition for a 2d Gaussian 
ideal chain in Sec.~II, and then we give the theoretical predictions
based on the blob picture in Sec.~III. In an experimental
setup, the escape transition is driven by changing the piston
separation $H$, while in a blob picture, the escape transition
is studied by changing the chain length $N$
or the strip length $L$ at fixed $H$ which is also the size of a blob.
Comparisons between the escape transition 
behavior of the Gaussian chain model and that of the blob model in the 
thermodynamic limit are given in Sec.~IV.
In Sec.~V we present our results from Monte Carlo (MC) simulations 
with the pruned-enriched Rosenbluth method (PERM)~\cite{g97} 
In Sec.~VI we provide a theoretical description based on the Landau
free energy approach which is compared with MC results.  
A summary and discussions are given in Sec.~VII. 
Detailed analyses of the variances of the imprisoned monomers are given in the Appendix.

\section{Escape for a 2d Gaussian chain}

For the escape transition of a Gaussian chain, 
a closed-form of the exact partition 
function was obtained earlier in ~\cite{ Klushin}. 
The asymptotic form of the free energy $F$ has two branches:
\be
F = \left\{
\begin{array}{ll}
N \frac{1}{2 d}\left(\frac{\pi a}{H}\right)^{2}=F_{\rm imp} & \enspace {\rm imprisoned \, state} \\
 \frac{\pi L}{H}=F_{\rm esc}  &\enspace {\rm escaped \, state}
\end{array} \right. \; .
\label{Fgauss}
\ee
Here $d$ is the dimensionality of space and a factor of $k_BT$ is absorbed in the free energy throughout the paper.
It was shown by direct numerical comparison that the simple asymptotic 
expressions provide a very accurate description. 
The two branches meet at the transition point which is given by
\be
\left(\frac{L}{Na}\right)^{**}=\frac{\pi}{2d}\frac{a}{H}\;, 
\enspace {\rm with} \enspace d=2  \; . \label{transgauss}
\ee 
The average lateral forces acting on the grafting point are obtained
by 
\be
f_L=\frac{\partial F(N,L,H)}{\partial L} \;   \label{fl}
\ee
which is related to the work required 
to pull a chain end into confined space by a unit distance.
The average compression force 
\be
f_H = - \frac{\partial F(N,L,H)}{\partial H} \label{fh}
\ee  
is related to the work of compression. It was shown earlier~\cite{Klushin} 
that the average fraction of imprisoned monomers is given by the 
derivative $N_{\rm imp}=\partial F(N,L,H)/\partial u$ of the free energy 
with respect to the effective confining potential 
$u=\left(\pi a/2 H\right)^{2}$. 

The average end-to-end distance per monomer is:
\be
 \frac{R}{Na} = \left\{
\begin{array}{ll}
N^{-1/2} & {\rm imprisoned \, state} \\
\frac{L}{Na} + N^{-1/2}\left(1-N_{\rm imp}/N\right)^{1/2} 
& {\rm escaped \, state}
\end{array} \right. \; .
\label{Rgauss}
\ee
The Landau order parameter $s$ for the escape transition was 
introduced in~\cite{Klushin}
as a stretching degree of the chain $r_N/N$ for the imprisoned state and 
as the stretching degree of the stem $L/n_{\rm imp}$ for the escaped state. 
Note that the order parameter is defined for an instantaneous configuration 
with a given end-to-end distance $r_N$ and the number of imprisoned monomers 
$n_{\rm imp}$ which may be quite different from the equilibrium average values 
$R=\left\langle r_N\right\rangle$ and 
$N_{\rm imp}=\left\langle n_{\rm imp}\right\rangle$. 
The Landau free energy is defined as a function of the order parameter. 
The Landau free energy for the imprisoned state is just the free energy 
of a coil as a function of its free end position, which has a standard 
parabolic form: 
\be
\Phi_{\rm imp}(s)=N\left[\frac{d}{2}s^2+\frac{1}{2 d}\left(\frac{\pi a}{H}\right)^2\right] \;.
\ee 
At $s=s_{\rm eq}^{\rm imp}=0$, the system is in equilibrium and 
the corresponding
equilibrium free energy, i.e., the depth of the minimum, is   
$\Phi_{\rm imp}(0)=(N/2d) (\pi a/H)^2$.
For the escaped state,
\be
\Phi_{\rm esc}(s)=\frac{L}{a}\left[\frac{d}{2}s+\frac{1}{2 d}
\left(\frac{\pi a}{H}\right)^2\frac{1}{s}
\right] \;.
\ee
The position of the minimum is at $s=s_{\rm eq}^{\rm esc}={\pi a}/{d H}$, 
and the corresponding depth is $\Phi_{\rm esc}(s_{\rm eq}^{\rm esc})={\pi L}/{H}$. 
The binodal is determined by the condition that the two minima 
are equally deep, 
i.e. $\Phi_{\rm imp}(s_{\rm eq}^{\rm imp})=\Phi_{\rm esc}(s_{\rm eq}^{\rm esc})$,
which leads to the transition point described by Eq.~(\ref{transgauss}). 
At the transition point, the average order parameter $S=<s>$ jumps
from $s_{\rm eq}^{\rm imp}$ to $s_{\rm eq}^{\rm esc}$.
The Landau function allows one to analyze metastable states and 
to define the two lines where either one of the metastable 
minima vanishes ~\cite{Skvortsov06}. 
Theoretical predictions for the escape transition of a Gaussian chain 
are summarized in Table~\ref{table1} and shown in Fig~\ref{fig-pre1}. 
Detailed discussion and comparison with the prediction from the blob picture 
for 2d polymer chains are given in Sec.~IV.

\section{Blob picture of a 2d escape}

A free chain in $d=2$, has an average end-to-end distance given by
($a$ is the distance between neighboring monomers) [17-20]
\be
     R_{\rm F}=aN^{3/4}    \label{Rfree}
\ee
here the prefactors of order unity are neglected throughout.
Based on the blob picture we have a cigar of blobs ($n_b$ blobs in total)
in the confined situation. Thus the average end-to-end distance is
\be
        R=n_b(2r_b)=n_bH          \label{Rh}
\ee
where $r_b$ is the blob radius. Within a blob, self-avoiding walks (SAW)
statistics holds, so if $g$ monomers belong to a blob
\be
       H=ag^{3/4}=2r_b\;, \enspace g=(H/a)^{4/3} \;.
\ee
Since every monomer of a chain in an imprisoned state
must be in a blob we furthermore have
\be
       N=gn_b=n_b (H/a)^{4/3}\;, \enspace n_b=N(H/a)^{-4/3} \;.
\ee
This yields, together with Eq.~(\ref{Rh}), the formula for the 
end-to-end distance
\be
        R/a=N(H/a)^{-1/3}          \label{Rb}
\ee
If $H$ is of the same order as $R_{\rm F}$, Eq.~(\ref{Rfree}),
one finds $R/a=N^{3/4}$, giving a smooth crossover to mushroom 
behavior, as expected.

   The free energy excess of the chain in an imprisoned
state (in units of $k_BT$), 
relative to an
unconfined mushroom, is simply the number of blobs, $n_b$
,
\be
   F_{\rm imp} =n_b =N (H/a)^{-4/3}     \label{Fimp}
\ee

We now define $N=N^*$ as the maximum chain length of an imprisoned
chain. Then for $N>N^*$ the chain consists of a ``stem" containing
$N^*$ imprisoned monomers and an escaped 
``crown" comprising the rest $N-N^*$ monomers (Fig.~1b).
Thus, we find $N^*$ from the condition that 
$R$ becomes equal to $L$ at the transition point, using Eq.~(\ref{Rb})
\be
         N^* =(L/a)(H/a)^{1/3} \;.   \label{ncr}
\ee
Since the free energy of an unconfined mushroom is taken as zero 
reference point, 
the ``crown" does not contribute to the excess free energy of the 
escaped chain, which is hence due to the stem only:
\be
     F_{\rm esc} = N^* (H/a)^{-4/3} =L/H \;.   \label{Fesc}
\ee
The average end-to-end distance of
the escaped chain (in axial direction parallel to the confining boundaries)
hence becomes
\be
     R_{\rm esc}= L+a(N-N^*)^{3/4} \;.
\ee
Equations~(\ref{Fimp}) and (\ref{Fesc}) show that the free energy 
as a function of $N$ for fixed $H$ and $L$, consists of two branches,
i.e., $F_{\rm imp}$ for the imprisoned state ($N<N^*$) and 
$F_{\rm esc}$ for the escaped state ($N>N^*$),
meeting at $N=N^*$.
The lateral and the compression forces are obtained from the free
energy by using Eqs.~(\ref{fl}) and (\ref{fh}). 
We use the same definition of the order parameter $s$ as that for the
Gaussian chain, so the average order parameter $S$
\be
S = <s>=\left\{
\begin{array}{ll}
 R/Na & \enspace {\rm imprisoned \, state} \\
 L/N^*a=(H/a)^{-1/3}  &\enspace {\rm escaped \, state}
\end{array} \right. \; . \label{sblob}
\ee
From Eqs.~(\ref{Rb}), (\ref{ncr}), and (\ref{sblob}), we find that the
order parameter does not show any
discontinuity at the transition, but simply stays constant,
i.e. $S=(H/a)^{-1/3}$.
Results of the theoretical predictions are listed in Table~\ref{table1}
and also shown in Fig.~\ref{fig-pre1}.

\section{Comparison of the Gaussian and Blob pictures}

\begin{figure*}
\begin{center}
\epsfig{file=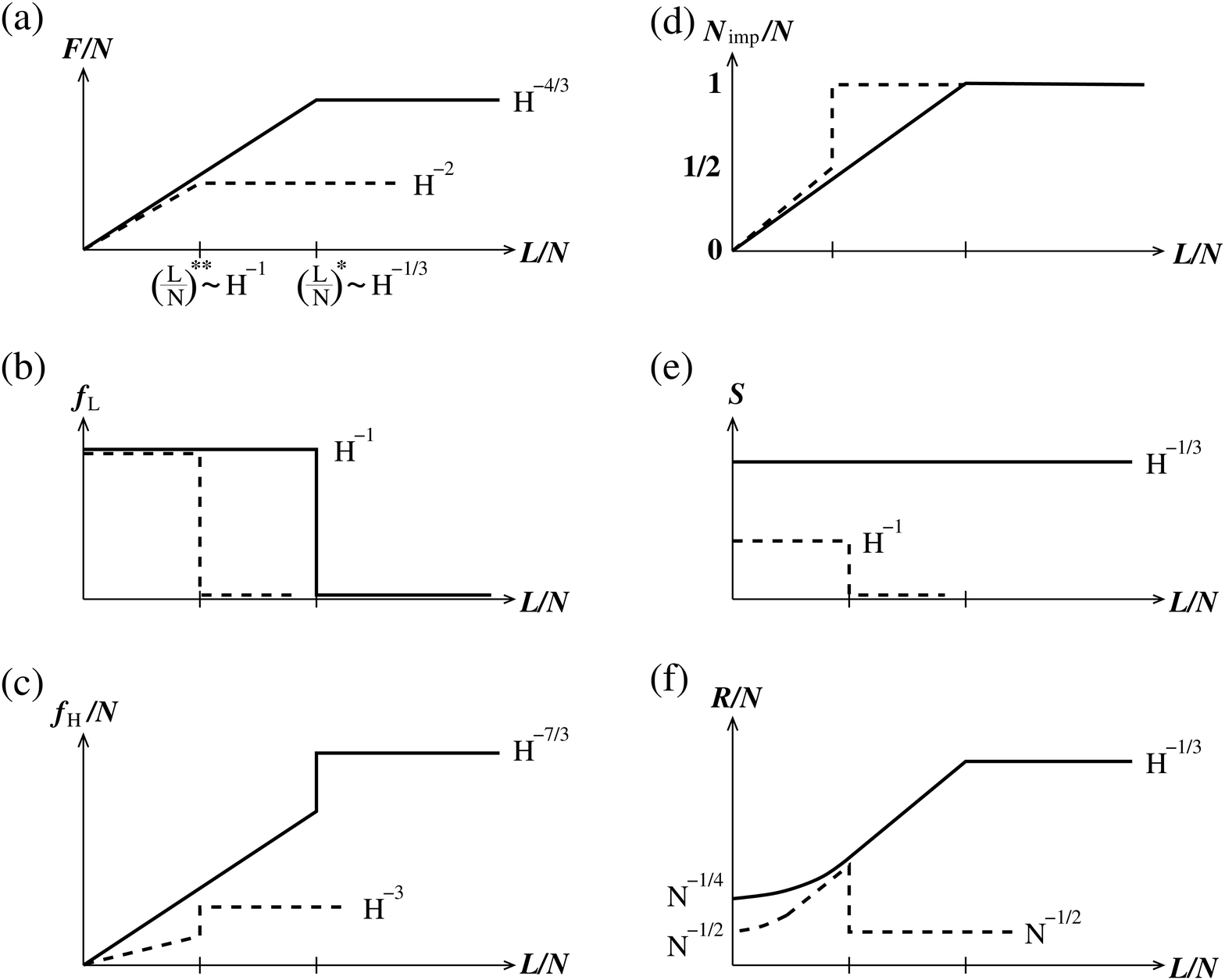, width=10.0cm, angle=0}
\caption{Theoretical predictions for various averaged chain characteristics
plotted against $L/N$ at constant strip width $H$: 
(a) the free energy per monomer $F/N$,
(b) the lateral force $f_L$,
(c) the compression force per monomer $f_H/N$,
(d) the fraction of 
imprisoned monomers $N_{\rm imp}/N$,
(e) the order parameter $S$,
(f) the end-to-end distance per monomer $R/N$, . Gaussian model results are shown by dotted lines, blob model results - by solid lines. The chain is in an imprisoned state
for $\frac{L}{N}>\left(\frac{L}{N}\right)^{**}$ 
$[\left(\frac{L}{N}\right)^{*}]$, and in
an escaped state for $\frac{L}{N}<\left(\frac{L}{N}\right)^{**}$
$[\left(\frac{L}{N}\right)^{*}]$ for the Gaussian chain model 
[blob picture].}
\label{fig-pre1}
\end{center}
\end{figure*}

\begin{table*}
\begin{center}
\caption{Theoretical predictions for the average values of the free energy 
per monomer $F/N$,
the lateral force $f_L$, the compression force per monomer $f_H/N$,
the fraction of imprisoned monomers $N_{\rm imp}/N$, 
the order parameter $S$, and the reduced end-to-end distance $R/Na$
based on the Gaussian chain model and the blob picture.}
\label{table1}
\begin{ruledtabular}
\begin{tabular}{|c|c|c|c|c|}
{Characteristics}   &  \multicolumn{2}{c|} {Imprisoned}  & 
\multicolumn{2}{c|} {Escaped} \\ 
 {of the chain}   &  \multicolumn{2}{c|} {state}  &
\multicolumn{2}{c|} {state} \\ \hline
& Gaussian & Blob & Gaussian & Blob  \\ \hline
$\frac{F}{N}$  & $\left(\frac{\pi a}{2H}\right)^2$ & $ \left(\frac{a}{H}\right)^{4/3}$ 
& $\frac{\pi}{H} \frac{L}{N}$ & $\frac{1}{H} \frac{L}{N}$ \\
$f_L$ & 0  & 0 & $\frac{\pi}{H}$ & $\frac{1}{H}$ \\
$\frac{f_H}{N}$ & $\frac{\pi^2}{2a}\left(\frac{a}{H}\right)^3$ &
$\frac{4}{3a}\left(\frac{a}{H}\right)^{7/3}$ & $\frac{\pi}{H^2} \frac{L}{N}$ &
$\frac{1}{H^2} \frac{L}{N}$ \\
$\frac{N_{\rm imp}}{N}$ & 1 & 1 & $\frac{2}{\pi}\frac{H}{a} \frac{L}{Na}$ &
$ \left(\frac{H}{a}\right)^{1/3} \frac{L}{Na}$ \\
$S$ & 0 & $\left(\frac{a}{H}\right)^{1/3} $ & $\frac{\pi}{2} \frac{a}{H}$ 
& $\left(\frac{a}{H}\right)^{1/3}$ \\
$\frac{R}{Na}$ & $N^{-1/2}$ & $\left(\frac{H}{a}\right)^{-1/3}$ &
$N^{-1/2}\left(1-\frac{N_{\rm imp}}{N}\right)^{1/2}+\frac{L}{Na}$ &
$N^{-1/4}\left(1-\frac{N_{\rm imp}}{N}\right)^{3/4}+\frac{L}{Na}$ \\
\end{tabular}
\end{ruledtabular}
\end{center}
\end{table*}

   In the thermodynamic limit $N\rightarrow \infty$, $L \rightarrow \infty$,
$L/N$ remains as a nontrivial variable along with $H$. In Fig.~\ref{fig-pre1}, 
theoretical predictions of the Gaussian chain model and of the blob model 
are shown by dotted and solid lines, respectively. The strip width, $H$, 
is fixed, and the ratio $L/N$ is varied.
The chain is in an imprisoned state if $\frac{L}{N}$ is larger than the 
corresponding critical value,  $\left(\frac{L}{N}\right)^{**}$ for the 
Gaussian chain model or $\left(\frac{L}{N}\right)^{*}$ for the blob model, 
and in the escaped state for for $\frac{L}{N}<\left(\frac{L}{N}\right)^{**}$
$[\left(\frac{L}{N}\right)^{*}]$. 
All curves show only the scaling behavior disregarding numerical 
coefficients of order one, and the bond length $a$ is taken as a 
unit length. 

In both models, the free energy per monomer is given by piecewise 
linear functions 
of $L/N$. For the escaped state, the slope of $F/N$ vs. $L/N$ is the same 
up to numerical coefficients of order 1, and scales as $H^{-1}$. This slope 
has the meaning of the lateral force acting on the grafting point. It is 
proportional to the inverse size of a blob which is defined purely 
by confinement width $H$ irrespective of whether excluded volume 
interactions are present or not. The free energy per monomer in 
the imprisoned state is 
independent of $L/N$, and scales as the inverse number of monomers in 
one blob, $g^{-1}=\left(a/H\right)^{1/\nu}$, where $\nu$ is the Flory 
exponent. The transition points in the two models may be quite far apart 
since they scale differently with $H$. In the setup where $L/N$ is fixed 
and the piston separation is decreasing, the blob model predicts the 
transition to happen at lower compression, as compared to the Gaussian chain.
In both models the lateral force jumps from $~H^{-1}$ to zero at the 
respective transition point. As for compression forces, they are strongly 
affected by excluded volume interactions (Fig.~\ref{fig-pre1}c). 
The difference in the plateau 
values for the imprisoned state reflects the lower compressibility of the 
self-avoiding walk as compared to the Gaussian chain. This corresponds 
directly to the plateau values of the free energy in Fig.~\ref{fig-pre1}a. In the 
Gaussian chain model, the compression force jumps at the transition point by 
a factor of 2, while the jump in the blob model is by a factor of 4/3.

Figures~\ref{fig-pre1}a - \ref{fig-pre1}c suggest that the behavior of both models is 
fundamentally the same and characteristic of a first-order phase 
transition. However, the next 3 graphs demonstrate qualitatively 
different predictions of the Gaussian and the blob models.

For the Gaussian chain model, a transition from the confined state to the 
escaped state is accompanied by a jump in the average number of imprisoned 
monomers. At the transition point, one half of the total number of monomers are ejected 
outside to form a crown, see Fig.~\ref{fig-pre1}d. In contrast to that, the blob model 
predicts a smooth change without a jump. This directly affects the behavior 
of the average order parameter: while in the Gaussian chain model there is a 
pronounced jump, the blob model suggests that the order parameter does not 
change at all. The large constant value of $S$ in the blob model is due to 
the cigar-shape conformation of the chain in an imprisoned state
which is identical to 
the shape of the stem in the escaped state.

The behavior of the average end-to-end distance is also qualitatively different 
for the two models. The plateau value for the imprisoned Gaussian chain, 
$R/N =N^{-1/2}$, is characteristic of the ideal coil, while for the 
elongated cigar, this ratio is $N$-independent. The linear part of the curves 
corresponding to the escaped state simply represents the dominant contribution
of the stem to the overall end-to-end distance, $R \approx L$ irrespective 
of the model. At very low values of $L/N$ corrections due to the crown size 
become large, as shown in Fig~\ref{fig-pre1}f.  
The size of the Gaussian chain demonstrates a jump at the transition point 
consistent with a strong conformation change accompanying the first-order 
transition. In the blob model, the chain size does not have any jump. 

It is clear that the predictions of the blob model presented in Fig.~\ref{fig-pre1} 
contain some internal contradictions. On the one hand, the picture of 
the two branches of the free energy meeting at some angle suggests a first 
order transition.  The jumps in the lateral and compression forces are 
a simple consequence of that. On the other hand, nothing dramatic happens 
to the chain conformation in the blob picture: the change from a completely 
confined state to a state with a small escaped tail is continuous, as 
evidenced in Figs~\ref{fig-pre1}e and ~\ref{fig-pre1}f. 
The presence of discontinuity in the slope 
of the fraction of imprisoned monomers suggests that the transition should be
classified as second order.

A crucial question to ask is whether one can identify two distinct
separate states with a bimodal distribution of some appropriate
order parameter. This we address by employing a Monte-Carlo 
simulation of $2d$ self-avoiding chains undergoing the escape
transition.

\section{Monte Carlo simulation}

\begin{figure}
\begin{center}
\epsfig{file=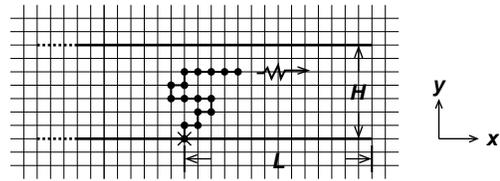, width=6.5cm, angle=0}
\caption{Schematic drawing of a polymer chain growing
as a self-avoiding walk inside a finite strip and grafted
at $(x=0,y=0)$. Monomers are allowed
to sit on the lattice sites except for the lattice sites representing the walls
 $\{-L\leq x \leq L\;,\,y=0\}$ and
$\{-L\leq x \leq L\;,\,y=H\}$.
The first monomer is attached with a bond to the grafting
site marked by a cross. Lengths are measured in units of the
lattice spacing.}
\label{fig-strip-esc}
\end{center}
\end{figure}

  Single polymer chains grafted in the middle of a
strip of length $2L$ and width $H$ are
described by SAWs of $N$ steps on a square lattice between
two hard walls with distance $H$ as shown in Fig.~\ref{fig-strip-esc}.
Monomers are supposed to sit on lattice sites but they are forbidden to
sit on the two walls, i.e. $\{-L\leq x \leq L\;,\, y=0\}$ and
$\{-L\leq x \leq L\;, \, y=H\}$.
For our simulations we employ PERM and PERM with $k$-step Markovian anticipation as described
in Ref.~\cite{g97,Hsu03}.
PERM is a biased chain growth algorithm
with population control. Polymer chains are built like random
walks by adding one monomer at each step. Thus it has the advantage
of estimating the partition sum and counting the imprisoned
monomers directly.

We simulate 2d SAWs starting at the grafting point of the strip
of length $L=800$, $L=1600$, $3200$, and $6400$.
The width of the strip is varied from $H=5$ to $H=129$. Depending
on the chosen sizes of $L$ and $H$,
the total chain length is varied from $2500$ to $50000$ in order
to cover the transition region.

\begin{figure}
\begin{center}
\epsfig{file=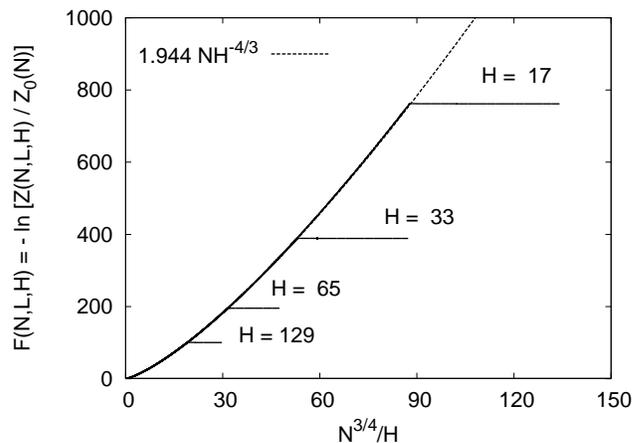, width=6.0cm, angle=270}
\caption{(a) Free energy relative to a free chain,
$F(N,L,H)=-\ln[(Z(N,L,H)/Z_0(N)]$, plotted against $N^{3/4}H^{-1}$ 
for $L=6400$ and $H=17$, 33, 65, and 129.
The dashed curve is $F_{\rm imp}(N,L,H)=1.944NH^{-4/3}$ and gives
the best fit of the data.} 
\label{fig-F-L6400}
\end{center}
\end{figure}

\begin{figure}
\begin{center}
\epsfig{file=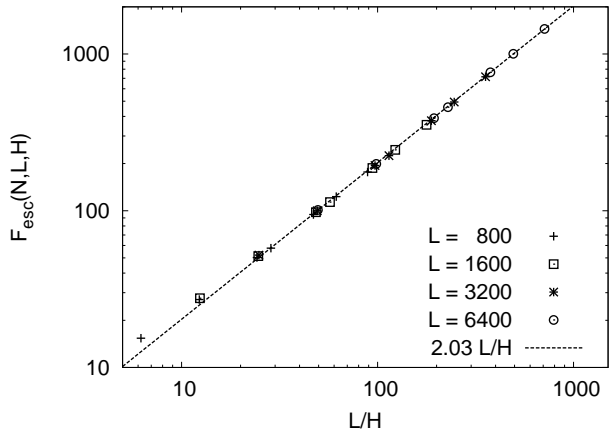, width=6.0cm, angle=270}
\caption{The log-log plot of the excess free energy of the escaped
chain, $F_{\rm esc}(N,L,H)$,
plotted against $L/H$ for various values of $L$ and $H$.
The dashed line is $F_{\rm esc}(N,L,H) = 2.03L/H$ and gives the 
best fit of the data.}
\label{fig-fesc}
\end{center}
\end{figure}

\begin{figure}
\begin{center}
\epsfig{file=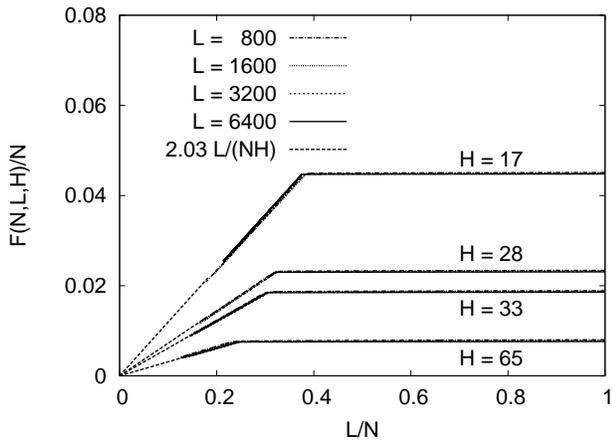, width=6.0cm, angle=270}
\caption{Free energy relative to a free chain divided by $N$, 
$F(N,L,H)/N$, plotted against $L/N$ for various values of 
$L$ and $H$. 
The dashed line extrapolated to zero is $2.03L/(NH)$.}
\label{fig-zall}
\end{center}
\end{figure}

\subsection{Free energy}                                                                               
Let us first discuss the scaling behavior of the free energy.
The partition sum of a free SAW in infinite volume for $N \rightarrow \infty$
scales as
\be
Z_0(N)=\mu_\infty^{-N}N^{\gamma-1}
\ee
with $\mu_\infty$ being the critical fugacity per monomer, and
with $\gamma=43/32$ being a universal exponent~\cite{Eisenriegler}.
As chains are still confined in a strip of width $H$, one should expect 
the scaling laws of the excess free energy including the crossover from 
the region of wide strips, $R_F<<H$, ($R_F \sim N^{\nu}$ is the Flory radius 
with $\nu=3/4$ in $d=2$),
to the region of narrow strips, $R_F>>H$, where chains are
stretched. Since the length of the strip
is finite, here we are more interested in another expected crossover 
behavior of the excess free energy from an imprisoned and stretched
chain state to an escaped state.
Therefore, we plot the excess free energy
$F(N,L,H)=-\ln(Z(N,L,H)/Z_0(N))$ against $N^{3/4}H^{-1}$ with the
precise estimate of $\mu_\infty=0.37905228$~\cite{Guttmann} 
in Fig.~\ref{fig-F-L6400}.
The partition sum $Z(N,L,H)$ is the total number of possible configurations 
of SAW of $N$ steps partially confined in a strip of length $2L$ and of 
width $H$, which is estimated directly in the simulation. 
In Fig.~\ref{fig-F-L6400}, the sharp crossover behavior from
the imprisoned states to the escaped states is indeed seen as 
$N$ increases for a fixed value of $H$. 
The excess free energy of the escaped chain is independent of $N$,
as predicted in a blob picture by Eqs.~(\ref{Fimp}) and (\ref{Fesc}).
The best fit of the free energy for imprisoned state is given by 
\be
    F_{\rm imp}(N,L,H) \sim 1.944(2) NH^{-4/3} \;.  \label{Fimps}
\ee
It is in perfect agreement with the previous estimation in
Ref.~\cite{Hsu03}, where the fugacity per monomer
scales as $\mu_H-\mu_\infty \sim 0.737H^{-4/3}$
and hence the free energy of the chains of size $N$ in 
the imprisoned state scales as
$F_{\rm imp} \sim \frac{0.737}{\mu_\infty}NH^{-4/3} \approx 1.944 NH^{-4/3}$.
Values of the excess free energy of the escaped chain, $F_{\rm esc}(N,L,H)$ 
are determined by the horizontal curves shown 
in Fig.~\ref{fig-F-L6400}. 
Results for $L=800$, $1600$, $3200$ and $6400$ are shown 
in Fig.~\ref{fig-fesc}, where we obtain 
\be
F_{\rm esc}(N,L,H)=2.03(3)L/H \;. \label{Fescs}
\ee
As $H$ becomes comparable to $L$, i.e. $L/H \sim O(10)$, the data points
deviate slightly from the straight line, indicating that there are further
finite size corrections for small $L/H$.
In order to compare with the theoretical prediction shown in 
Fig.~\ref{fig-pre1}a, we plot $F(N,L,H)/N$ against $L/N$ for
various values of $L$ and $H$ in Fig.~\ref{fig-zall}. 
In the escaped regime, these straight lines extrapolated to $L/N=0$
are indeed described by Eq.~(\ref{Fescs}) very well.
With conventional Monte Carlo simulations it is difficult to estimate the
partition sum precisely.
With PERM we do have very precise estimates of $Z(N,L,H)$, and therefore
we can obtain the lateral and the compression forces by differentiating
the estimated free energy, Eqs.~(\ref{Fimps}) and (\ref{Fescs}), with respect 
to $L$ and $H$ respectively.  

\begin{figure}
\begin{center}
\epsfig{file=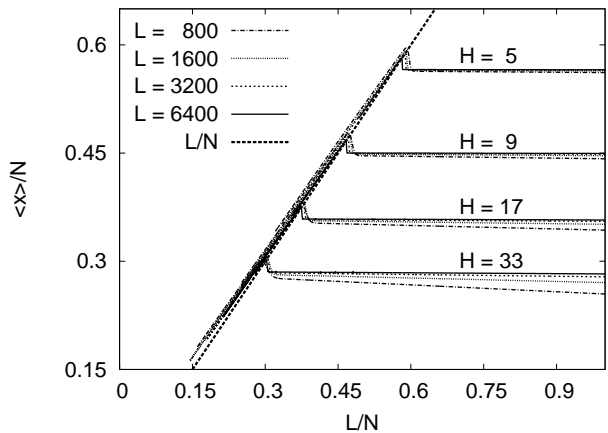, width=6.0cm, angle=270}
\caption{Average end-to-end distance divided by $N$, $<x>/N$,
plotted against $L/N$ for various values of $L$ and $H$.
The dashed line is $<x>/N=L/N$.}
\label{fig-x}
\end{center}
\end{figure}

\begin{figure}
\begin{center}
\epsfig{file=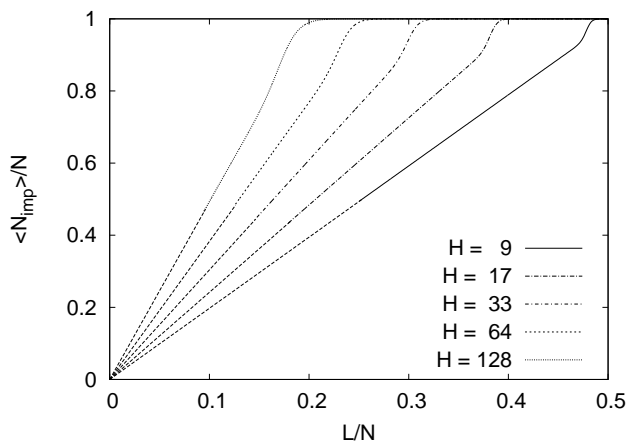, width=6.0cm, angle=270}
\caption{Average fraction of the imprisoned monomers, $<N_{\rm imp}>/N$,
plotted against $L/N$ for $L=800$ and various values of $H$.
The jump becomes more prominent as $H$ decreases.
The ends of the extrapolated straight dashed lines to $L/N=0$ all
approach zero.}
\label{fig-m}
\end{center}
\end{figure}
                                                                                
\begin{figure}
\begin{center}
\epsfig{file=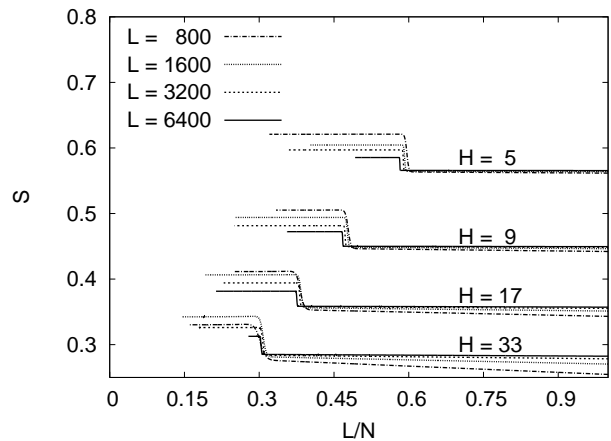, width=6.0cm, angle=270}
\caption{Average order parameter $S$
plotted against $L/N$ for various values of $L$ and $H$.}
\label{fig-s}
\end{center}
\end{figure}

\subsection{Average characteristics of polymer chains}

The most straightforward way to understand how the conformations
of the polymer chains change as they undergo the escape transition
is to estimate the end-to-end distance, $<x>$. 
In Fig.~\ref{fig-x}, we plot the end-to-end distance per monomer
$<x>/N$ versus $L/N$. As long as the polymer chains are imprisoned 
the curves are horizontal, i.e. the degree of chain stretching is constant, and 
$<x>$ increases linearly with $N$ as shown in 
Ref.~\cite{Hsu03}. As the width of the strip, $H$, decreases,
the chains are stretched more.
As $L/N$ decreases, we see that there is a jump in each curve to another branch where
$<x>=L$ This means that the chain stretching is abruptly increased so that 
the chain reaches the edge of the strip, indicating that the transition to a partially escaped conformation is first-order like.
This is in contrast to the smooth behavior predicted by the blob picture and
shown in Fig.~\ref{fig-pre1}f.

A jump-wise change in the chain stretching suggests a similar change in the
average number of imprisoned monomers, $<N_{\rm imp}>$.
In Fig.~\ref{fig-m}, we plot the fraction of imprisoned monomers
$<N_{\rm imp}>/N$ versus $L/N$ for $L=800$. As long as the chain is 
imprisoned, $<N_{\rm imp}>/N=1$.
With the increase in the number of monomers, $N$,
each curve indeed develops a jump and the jump becomes more 
pronounced for smaller $H$ .
For a fixed $L$ and  $N\rightarrow \infty$, $<N_{\rm imp}>/N \rightarrow 0$ in accord with both theoretical models.

In Fig.~\ref{fig-s}, we plot the average of the order parameter $S$,  
versus $L/N$. We see clear jumps between the
two states.
For a given value of $H$, the order parameter 
behaves as a step function. 
The data seem to suggest that the magnitude of the jump decreases 
with the size of the system, $L$. A detailed analysis of the
distributions of $s$ will be presented in the next section.
We will see that the apparent decrease in the jump is due to poor sampling 
of the escaped state. 

\subsection{Transition points}

For our simulations the transition points can be determined by analyzing three 
quantities: (1) free energy, (2) variance of the number of imprisoned monomers 
$N_{\rm imp}$, and (3) variance of the end-to-end distance $x$. 
Since at the transition point the free energy of the imprisoned state
is equal to the free energy of the escaped state,
Eqs.~(\ref{Fimps}) and (\ref{Fescs}) give the following relation between
$L$, $N$ and $H$ at the transition point:
\be
       \left(\frac{N}{L}\right)_{\rm tr}=1.04(2)H^{1/3} \; .  \label{f-nc-e}
\ee
This shows that a polymer chain of size $N$ can be confined in a strip by
tuning the length $L$ or the width $H$ of the strip.

\begin{figure}
\begin{center}
\epsfig{file=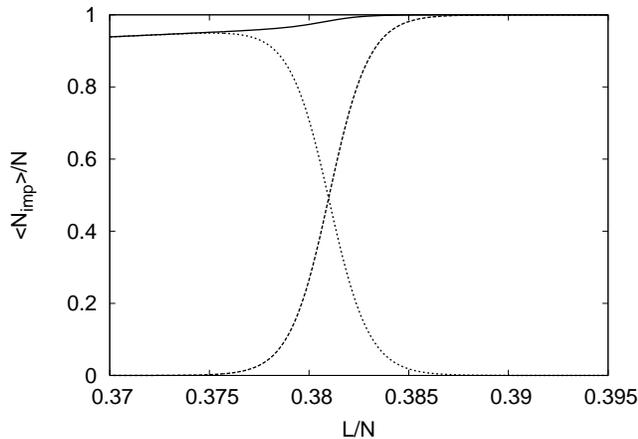, width=6.0cm, angle=270}
\caption{Average fraction of the imprisoned monomers $<N_{\rm imp}>/N$ 
(solid line) and two partial contributions, $<N_{\rm imp}>_1/N$ due to 
confined configurations, and $<N_{\rm imp}>_2/N$ due to escaped configurations 
(dotted lines) as functions of $L/N$ near the transition point 
$(L/N)_{\rm tr} = 0.381$ for $L=3200$ and  $H=17$. }
\label{fig-m12}
\end{center}
\end{figure}

\begin{figure*}
\begin{center}
$\begin{array}{c@{\hspace{0.1in}}c}
\multicolumn{1}{l}{\mbox{\large (a)}} &
        \multicolumn{1}{l}{\mbox{\large (b)}} \\ [-0.53cm]\\
\epsfig{file=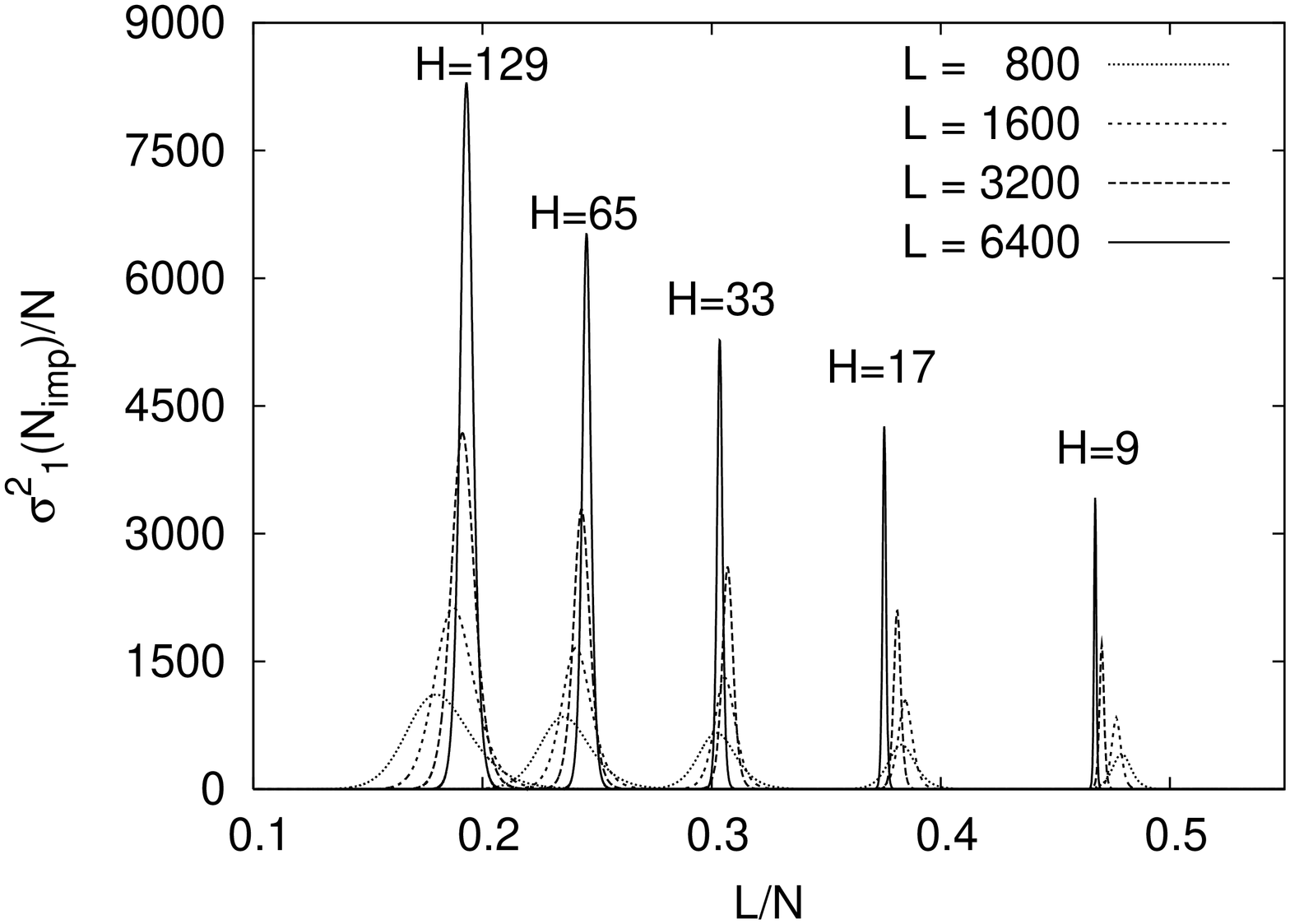, width=6.0cm, angle=270} &
\epsfig{file=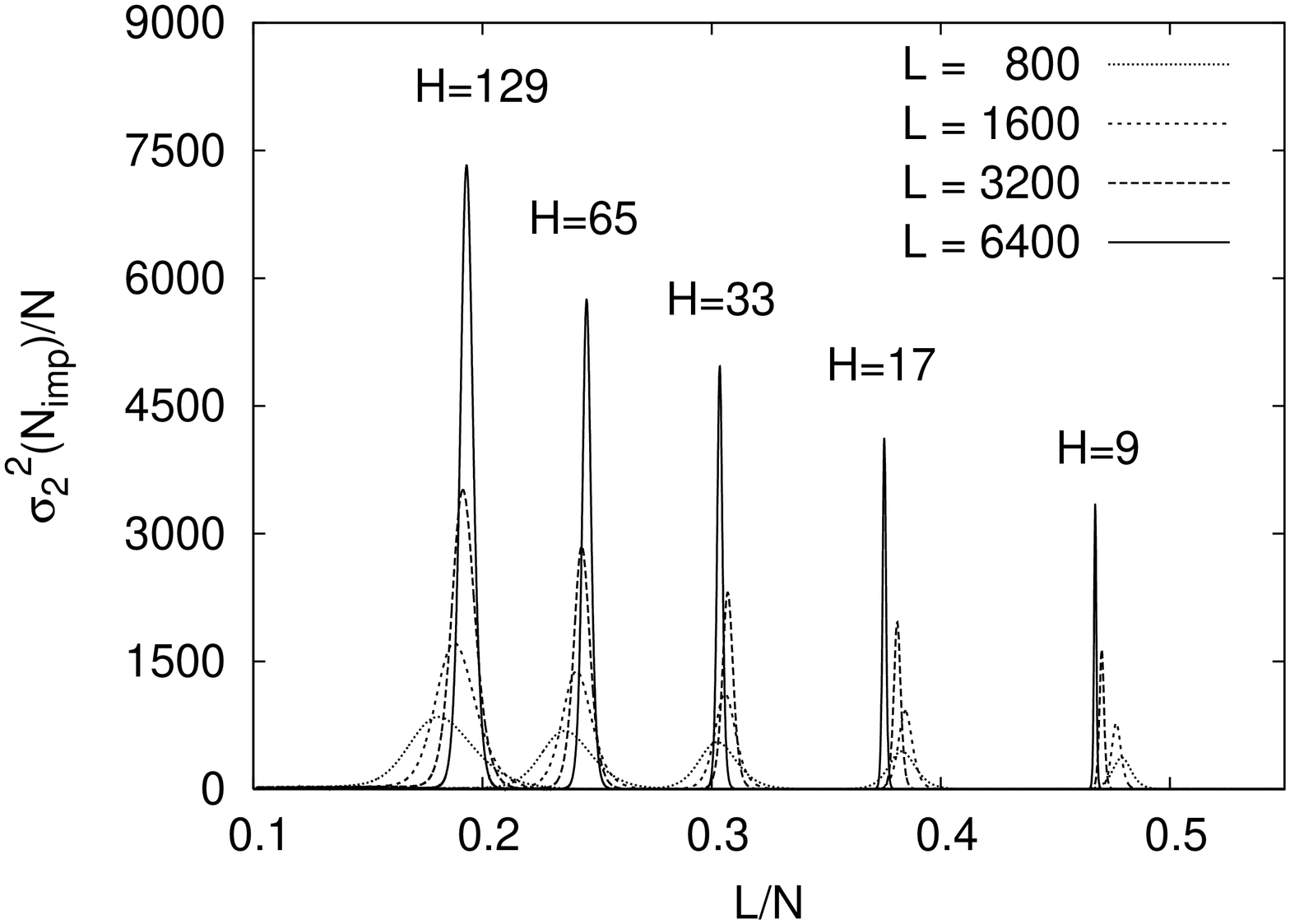, width=6.0cm, angle=270} \\
[0.4cm]
\end{array}$
\caption{Variances of the number of imprisoned monomers divided by $N$, (a)
$\sigma_1^2(N_{\rm imp})/N$, for the imprisoned state, and
(b) $\sigma_2^2(N_{\rm imp})/N$, for the escaped state,
plotted against $L/N$.
The height of peaks increases with $L$ for a fixed
value of $H$.}
\label{fig-m2}
\end{center}
\end{figure*}

The abrupt change scenarios of $<x>/N$ and $<N_{\rm imp}>/N$
shown in Fig.~\ref{fig-x} and \ref{fig-m} indicate a phase transition 
but it is difficult to locate precisely the transition point.

It is clear that all the chain configurations can be divided into two subsets:
imprisoned and escaped. Far from the transition point, only one subset 
is important in defining the average characteristics, 
but in the vicinity of the transition point both subsets contribute, 
as shown in Fig.~\ref{fig-m12}. 
The average $<N_{\rm imp}>$ can be rewritten as
\ba
     <N_{\rm imp}>&=&\frac{\sum_{\left\{C_1\right\}}N_{\rm imp} W_N(C_1)+
\sum_{\left\{C_2\right\}}N_{\rm imp} W_N(C_2)}
{\sum_{\left\{C_1\right\}}W_N(C_1)+\sum_{\left\{C_2\right\}}W_N(C_2)}
\nonumber \\
  &=& <N_{\rm imp}>_1+<N_{\rm imp}>_2 \label{Nimpa}
\ea
where $C_1$ $(C_2)$ denotes the imprisoned (escaped) configurations,
$W_N(C_1)$ $(W_N(C_2))$ are the total weights of the chain for obtaining the
configuration $C_1$ ($C_2$), and $<\ldots>_\alpha$ denotes the 
partial contributions due to the imprisoned configurations ($\alpha=1$) or 
the escaped configuration ($\alpha=2$).
Similarly,  
\be
     <N_{\rm imp}^2>= <N_{\rm imp}^2>_1+<N_{\rm imp}^2>_2 \; .\label{Nimpa2}
\ee
Finally, the variance of $N_{\rm imp}$, 
$\sigma^2 (N_{\rm imp}) = <N_{\rm imp}^2>-<N_{\rm imp}>^2$ 
can be expressed as follows,
\be
    \sigma^2 (N_{\rm imp}) = \sigma_1^2(N_{\rm imp}) + 
\sigma_2^2(N_{\rm imp})-2<N_{\rm imp}>_1<N_{\rm imp}>_2
\ee
where $\sigma_{1,2}^2(N_{\rm imp})=<N_{\rm imp}^2>_{1,2}-
<N_{\rm imp}>_{1,2}^2$. 
It is shown in the Appendix that the variances of partial contributions 
to the total number of imprisoned monomers, $\sigma_1^2(N_{\rm imp})$ and 
$\sigma_2^2(N_{\rm imp})$ are much better suited for locating the  
transition point than the full variance $\sigma^2 (N_{\rm imp})$ 
because of the very asymmetric behavior of the latter. 
We present $\sigma_1^2(N_{\rm imp})/N$ and $\sigma_2^2(N_{\rm imp})/N$ as functions 
of $L/N$ in Fig.~\ref{fig-m2} for various values of $H$ and $L$. 
It is clear that for a fixed width $H$, the peaks become
sharper as the length $L$ of the strip increases.
The escape transition points are identified with the positions of
the peaks of $\sigma_1^2(N_{\rm imp})$ and $\sigma_2^2(N_{\rm imp})$ as determined by
a curve fitting. 
Values of the transition points, $(L/N)_{\rm tr , 1}$ and 
$(L/N)_{\rm tr , 2}$ are the same
to the third digit for fixed values of $L$ and $H$, so the transition point
is taken as an average of them, 
i.e. $(L/N)_{\rm tr}=((L/N)_{\rm tr , 1}+(L/N)_{\rm tr , 2})/2$.
Results of $(L/N)_{\rm tr}$ obtained by this method are listed in 
Table~\ref{table2} and presented in Fig.~\ref{fig-nc-new}.
A more detailed discussion are given in the Appendix.
A similar method was also used for determining the transition point from
the variance of the end-to-end distance $x$.
Results are also listed
in Table~\ref{table2}and presented in Fig.~\ref{fig-nc-new}.
All the values of $(N/L)_{\rm tr}$ are plotted in Fig.~\ref{fig-nc-new}
against $H^{1/3}$.  The best fit gives
\be
    \left(\frac{N}{L}\right)_{\rm tr}=1.025(35)H^{1/3}    \label{f-nc-new-e}
\ee
which is consistent with the estimation of Eq.~(\ref{f-nc-e}) within the error bar.

\begin{table*}
\begin{center}
\caption{Values of the transition points,
$(L/N)_{\rm tr}$, determined from the analysis of the variances
$\sigma_1^2(N_{\rm imp})$ and $\sigma_1^2(x)$ for the imprisoned states,
and from the variances $\sigma_2^2(N_{\rm imp})$ and $\sigma_2^2(x)$
for the escaped states.}
\label{table2}
\begin{ruledtabular}
\begin{tabular}{r|rrrr|rrrr}
  &  \multicolumn{4}{c|} {$(L/N)_{{\rm tr}, N_{\rm imp}}$}  &
\multicolumn{4}{c} {$(L/N)_{{\rm tr}, x}$} \\ \hline
$H$ & $L=800$ & $1600$ & $3200$ & $6400$ & $L=800$ & $1600$ & $3200$ & $6400$\\ \hline
9   &  0.4790 & 0.4766 & 0.4703 & 0.4649 & 0.4790 & 0.4766 & 0.4723 & 0.4674 \\
17  &  0.3831 & 0.3846 & 0.3810 & 0.3754 & 0.3829 & 0.3836 & 0.3810 & 0.3754 \\
33  &  0.3027 & 0.3060 & 0.3071 & 0.3036 & 0.3024 & 0.3078 & 0.3070 & 0.3036 \\
65  &  0.2363 & 0.2412 & 0.2434 & 0.2455 & 0.2357 & 0.2409 & 0.2433 & 0.2455 \\
129 &  0.1816 & 0.1880 & 0.1916 & 0.1932 & 0.1803 & 0.1877 & 0.1914 & 0.1930 \\
 \end{tabular}
 \end{ruledtabular}
 \end{center}
 \end{table*}

\begin{figure}
\begin{center}
\epsfig{file=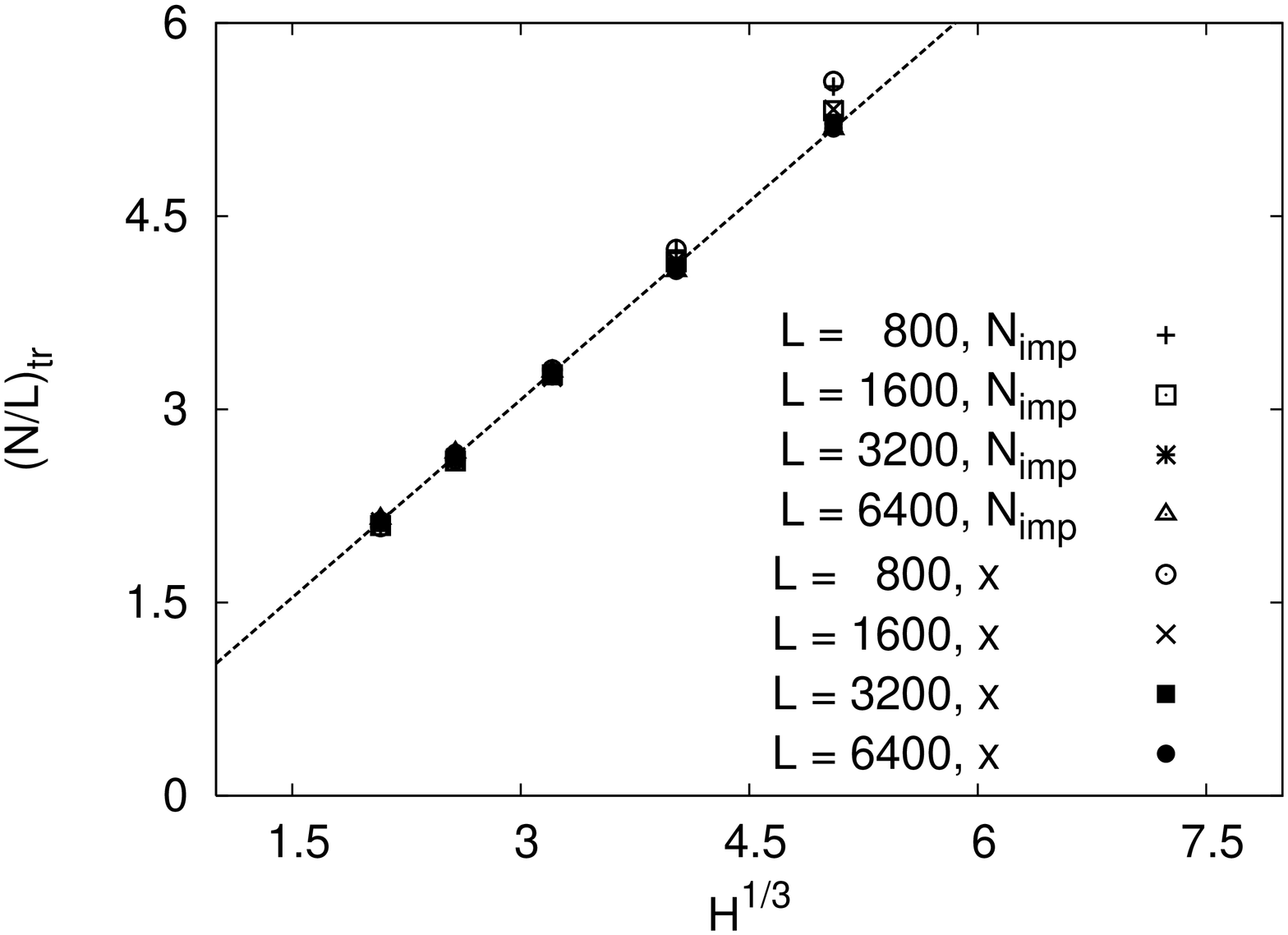, width=6.0cm, angle=270}
\caption{Transition points $(N/L)_{\rm tr}$ versus $H$.
The dashed line is $(N/L)_{\rm tr}=1.025(35)H^{1/3}$ and gives the
best fit of the data. }
\label{fig-nc-new}
\end{center}
\end{figure}

\section{Landau theory: Distribution of the order parameter} 

\subsection{Analytical theory}

The approach based on the Landau free energy is perfectly suited for
analyzing the relevant states in the escape problem, including
the metastable states. In the Landau theory, all the configurations
are first subdivided into subsets associated with a given value of
the order parameter, $s$ , and summation is performed separately 
within each subset. The full partition function can be obtained then
by integrating over the order parameter:
\be 
     Z=\exp(-F)=\int ds \exp(-\Phi(s))	\label{LFE}
\ee
where $\Phi(s)$ is the Landau free energy function,
i.e. the non-equilibrium
free energy taken as a function of the order parameter. In the 
vicinity of the first order transition point, the Landau free
energy is expected to have two minima (one stable and the other 
metastable). Our analysis will be based on finding the metastable 
minima and the associated thermodynamic characteristics. The proper
choice of the order parameter is not always obvious, nor are there
any standard recipes for making it. One criterion is quite clear:
the average value of the order parameter should allow one
to distinguish between two phases. For a first-order transition,
the average order parameter changes jump-wise. We require that
the properly chosen order parameter changes continuously
as the system evolves from a metastable state, through the 
transition state at the top of the barrier, and eventually falls 
into the equilibrium minimum. We have shown earlier~\cite{Skvortsov02}
that these criteria are satisfied if the order parameter is defined
as the chain stretching in the confined coil state, $s=r/(Na)$
where $r$ is the instantaneous end-to-end distance of the chain of $N$
monomers,
or as the stretching of the stem only in the flower conformation:
$s=L/(na)$, where $n$ is the number of 
monomers in the stem.
In analogy to the Gaussian case,
the Landau function consists of two branches that 
have to be introduced separately. As the chain is
in an imprisoned state, the Landau free energy is directly 
expressed in terms of the distribution of the end-to-end distance. 
There exists no closed-form formula for such a distribution of 
confined chains with excluded volume interactions. 
However, the distribution of the gyration radius 
for 3d chains confined in a tube was studied analytically and 
numerically in ~\cite{Victor2000}. 
It was proposed that the free energy of a confined chain with a given 
gyration radius $r_{g}$ 
can be presented as a sum of two terms:
\be
F(r_{g}) = N\left[A c^{\alpha}+B\left(\frac{r_{g}}{Na}\right)^{\delta}\right]\;,
\label{Frg}
\ee
where $c$ is the segment volume concentration expressed as a function of the 
gyration radius and the confinement geometry, $\alpha$ and $\delta$ are linked 
to the space dimension $d$ and the Flory exponent $\nu$ by 
$\alpha=\left(\nu d-1\right)^{-1}$ and $\delta=\left(1-\nu\right)^{-1}$. 
The first term describes the concentration effects in the des Cloizeaux~\cite{Cloizeaux} form, 
the second term is the Pincus~\cite{Pincus} scaling form of the stretching free energy, 
and A and B are model-dependent numerical coefficients of order unity.
Instead of $r_g$, we use the same ansatz, Eq.~(\ref{Frg}), to describe 
the end-to-end distance distribution by taking $c={Na^2}/{rH}$, $\alpha=2$,
and $\delta=4$. The free energy of the chain in an imprisoned state 
as a function of $s$
is hence given by 
\be
   \Phi_{\rm imp}(s) = N\left[A \left(\frac{a}{sH}\right)^2+Bs^4\right]\;,
\enspace s \leq \frac{L}{Na}  \, .\label{Pimp}
\ee
Since we prefer to keep the basic scaling formula of
the Landau free energy in order to provide predictions in a
simple analytical form, here we are not going to consider the further logarithmic 
correction terms as shown in ~\cite{Victor2000}.   

In the thermodynamic limit, the average value of 
the order parameter for the imprisoned state, $S_{\rm imp}$,
is found by locating the minimum of $\Phi_{\rm imp}(s)$,   
i.e. $d\Phi_{\rm imp}(s)/ds=0$ at $s=s_{\rm eq}^{\rm imp}$, and hence
\be
     S_{\rm imp}=s_{\rm eq}^{\rm imp}=(A/2B)^{1/6}(a/H)^{1/3} \;.    \label{Simp}
\ee
The minimum of the Landau free energy gives the free 
energy for the imprisoned state at equilibrium
\be
  F_{\rm imp}=\Phi_{\rm imp}(S_{\rm imp})=
   3B \left(\frac{A}{2B}\right)^{2/3}
  \left(\frac{a}{H}\right)^{4/3}N\;.           \label{Fimpl}
\ee
Compared with Eq.~(\ref{Fimp}), this is indeed the correct
scaling of the free energy.
The end-to-end distance at equilibrium is found as 
\be
    R_N=NaS_{\rm imp}=(A/2B)^{1/6} (H/a)^{-1/3}Na \;,       \label{Rimp}
\ee
which is consistent with the result of the blob model, Eq.~(\ref{Rb}).

As the chain is in an escaped state, the formula of the free energy function is 
identical to Eq.~(\ref{Pimp}), but corrected for the fact that only the $n$
monomers that are part of the stem contribute:
\ba
  \Phi_{\rm esc}(s)&=&n\left[A \left(\frac{a}{sH}\right)^2+Bs^4\right] \nonumber \\
&=& \frac{L}{a} \left[A \left(\frac{a}{H}\right)^2 s^{-3} +Bs^3\right]\;,
\enspace s \geq \frac{L}{Na}  \;.     \label{Pesc}
\ea
The average value of the order parameter in the escaped state, $S_{\rm esc}$,
is found by locating the minimum of $\Phi_{\rm esc}(s)$ and is given by 
\be
     S_{\rm esc} =s_{\rm eq}^{\rm imp}=(A/B)^{1/6}(a/H)^{1/3}      \label{Sesc}
\ee
Thus, the free energy of the escaped chain at equilibrium is
\be
  F_{\rm esc}=\Phi_{\rm esc}(S_{\rm esc})=2(AB)^{1/2} \left(\frac{L}{H}\right) 
\label{Fescl}
\ee
The transition point is found from the condition that the two minima of 
the Landau free energy function are of equal depth. 
Using Eqs.~(\ref{Fimpl}) and ~(\ref{Fescl}) we get
\be
  \left(\frac{L}{Na}\right)^{*}=
\frac{3 }{2^{5/3}}\left(\frac{A}{B}\right)^{1/6}
\left(\frac{a}{H}\right)^{1/3}
\label{LN}
\ee

\begin{figure}
\begin{center}
\epsfig{file=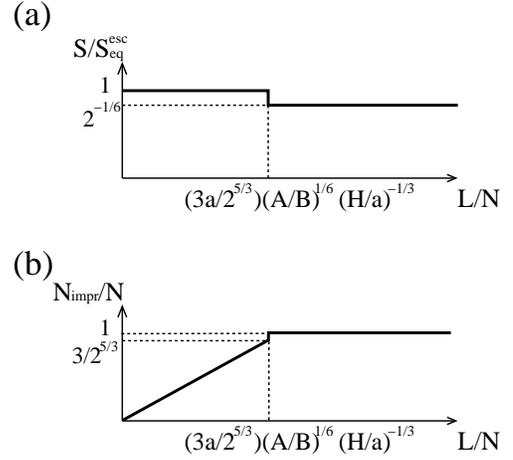, width=6.5cm, angle=0}
\caption{Based on the Landau theory, the theoretical predictions of
the average values of 
(a) the fraction of imprisoned monomers $N_{\rm imp}/N$,
and (b) the order parameter $S$,
are plotted against $L/N$. The chain is in an imprisoned state for
$L/N > (3a/2^{5/3})(A/B)^{1/6}(a/H)^{1/3}$, and in an escaped state
for $L/N < (3a/2^{5/3})(A/B)^{1/6}(a/H)^{1/3}$.}
\label{fig-Landau-pre}
\end{center}
\end{figure}

It is interesting to calculate the size of jumps implied by the Landau theory 
in the order parameter, the imprisoned monomers
and the end-to-end distance at the transition.
Using Eqs.~(\ref{Simp}) and (\ref{Sesc}) we immediately get the reduced jump of the 
order parameter
\be
\frac{\Delta S}{S_{\rm esc}}=\frac{S_{\rm esc}-S_{\rm imp}}{S_{\rm esc}}
=1-2^{-1/6}\approx 0.1091 \label{deltas} \;, \label{redS}
\ee
which is independent of $H$ and the coefficients 
$A$ and $B$. 
For an imprisoned state,  $<N_{\rm imp}>=N$ by definition, 
while for the coexisting escaped state with
the same choice of $H$, $L$, and $N$ we have only 
$<N_{\rm imp}>=L/S_{\rm esc}$ monomers. From Eqs.~(\ref{Sesc}) and (\ref{LN}),
we obtain the relative reduction in the number of imprisoned monomers
\be
   \frac{\Delta N_{\rm imp}}{N}=\frac{N-L/S_{\rm esc}}{N}=
1-\frac{3}{2^{5/3}}\approx 0.055 \;. \label{redN}
\ee
This number has a simple meaning of the fraction of the chain escaping out of
the confinement at the transition point. It is much smaller than $1/2$ in the 
Gaussian chain model, but non-zero in contrast to the blob model.
Finally, the reduced jump of the end-to-end distance is obtained by 
combining Eqs.~(\ref{Rimp}) and (\ref{LN})  
\be
   \frac{\Delta R}{L}=\frac{L-R_N}{L}=1-\frac{2^{3/2}}{3}\approx 0.0572 \;.
\label{redR}
\ee
Eqs.~(\ref{redS})-(\ref{redR}) show that the sizes of jumps in $S$, 
$<N_{\rm imp}>/N$ and $R_N/L$ are universal quantities.
Results for the average order parameter $S$ and the average fraction 
of imprisoned monomers 
$<N_{\rm imp}>/N$ predicted by the Landau theory are shown in Fig.\ref{fig-Landau-pre}.
Comparing with the numerical results shown in Fig.~\ref{fig-m} and Fig.\ref{fig-s},
we see that the Landau theory a good qualitative agreement.

The predicted free energy of the chain at equilibrium, 
Eqs.~(\ref{Fimpl}) and (\ref{Fescl}) 
follow the same scaling behavior as obtained by the MC simulations shown
in Eqs.~(\ref{Fimps}) and ({\ref{Fescs}). This allows us to identify
the numerical values of the constants
$A$ and $B$ for our model: $A\approx 1.057$ and 
$B \approx 0.975$.

\subsection{Numerical comparisons}

Here we focus on the results of the Landau free energy
of polymer chains partially confined in a strip of width $H=28$
and of length $L=800$, $1600$, $3200$, and $6400$.
Since PERM gives the possibility to estimate directly the partition sum
and the properly normalized histograms, the Landau free energy as a
function of $s$, $\Phi(N,L,H,s)$, is given by
\be
 \Phi(N,L,H,s)= -\ln \left(\frac{P(N,L,H,s)}{Z_0(N)}\right)  \label{Pps}
\ee
where $P(N,L,H,s)=\sum_{walks} \delta_{s,s'}$ is the histogram of $s$, 
and the partition sum of the partially confined chains can be written as 
\be
        Z(N,L,H)=\sum_{s}P(N,L,H,s)   \label{ps}
\ee
in accordance with Eq.~(\ref{LFE}).
In Fig.~\ref{fig-zs-org}, we plot four sets of results of the Landau 
free energy per monomer $\Phi(N,L,H,s)/N$ versus the order parameter $s$
for $L=800$, $1600$, $3200$ and $6400$.
Since the transition point is near $H^{1/3}$, the histograms are
obtained for $N/L=3.05$, $3.10$ and $3.15$ for each set.
The predicted analytical results of 
$\Phi^P(s)=\Phi_{\rm imp}(s)$ for the imprisoned state and 
$\Phi^P(s)=\Phi_{\rm esc}(s)$ for the escaped state, given by Eqs.~(\ref{Pimp})
and (\ref{Pesc}) are also shown for comparison.    
On the left-hand side of the branch points, due to the
finite-size effect, we see that the excess free energy for the 
imprisoned state (the minimum of the curve) at $s=s_{{\rm eq},L}^{\rm imp}$
converges to the predicted value (the minimum of the curve 
$\Phi^P(s)$) of polymer chains 
confined in an infinite strip at $s=s_{\rm eq}^{\rm imp}$
slowly as $L$ increases but $s_{{\rm eq},L}^{\rm imp}$ is slightly
larger than $s_{\rm eq}^{\rm imp}$ as $L \rightarrow \infty$.
The difference between those curves 
corresponding to the different ratio $N/L$ is almost
invisible for a fixed value of $L$ as predicted by Eq.~(\ref{Pimp}).
On the right-hand side of the branch points, we see that only those curves
for $L=800$ finally develop a parabola-like behavior with 
fluctuations and they are more concave than those curves predicted
by Eq.~(\ref{Pesc}). 
It shows that PERM has difficulties to 
sample configurations in the escaped regime as $L$ increases
and gives an explanation why we should not trust the size of 
those jumps that appear in Fig.~\ref{fig-s} too much.
However, one can easily overlook the existence of
two minima in such a delicate situation.
With PERM, at least we are able to give evidence for this two minimum 
picture of the first-order like transition.
We also see that additional finite-size correction terms
should be taken into account for the theoretical predictions
in Eqs.~(\ref{Pimp}) and (\ref{Pesc}).    

Taking the results for $L=6400$ as a reference, 
we plot the same data but shift all other curves by some constants, 
$c_{0,L}=-0.00235$, $-0.00109$, $-0.00044$
for $L=800$, $1600$, and $3200$ to make the three branch points
for $N/L=3.05$, $3.10$, and $3.15$ coincide with each other in 
Fig.~\ref{fig-zs-fit}.
According to the prediction by Eq.~(\ref{Pimp}), we should expect
that the four curves for different values of $L$ overlap with each other
in the imprisoned regime.
In fact, it is not the case but the difference between these curves decreases
as $L$ increases, and finally they will converge to one curve as $L$ becomes
very large. In the escaped regime, surprisingly, we see that 
those curves corresponding to
different $L$ all overlap with each other for a fixed ratio of $N/L$
as predicted by Eq.~(\ref{Pesc}). Although the lack of data for larger
$L$, precludes very strong conclusions, we may assume that 
these curves all show the same behavior as the curve for
$L=800$, and do further analysis. 

\begin{figure}
\begin{center}
\epsfig{file=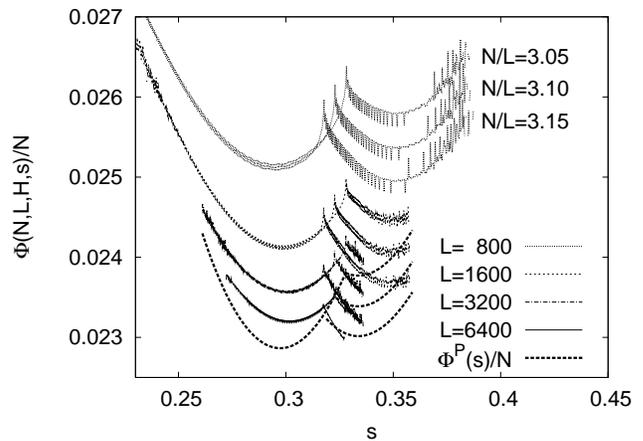, width=6.0cm, angle=270}
\caption{The Landau free energy divided by $N$, 
$\Phi(N,L,H,s)/N$, plotted
against $s$ for various values of $L$ and $H=28$. 
The predicted Landau free energy functions, 
$\Phi^P(s)=\Phi_{\rm imp}$, Eq.~(\ref{Pimp}), in the imprisoned regime
and $\Phi^P(s)=\Phi_{\rm esc}$, Eq.~(\ref{Pesc}), in the escaped regime
are also plotted (dashed lines).}
\label{fig-zs-org}
\end{center}
\end{figure}

\begin{figure}
\begin{center}
\epsfig{file=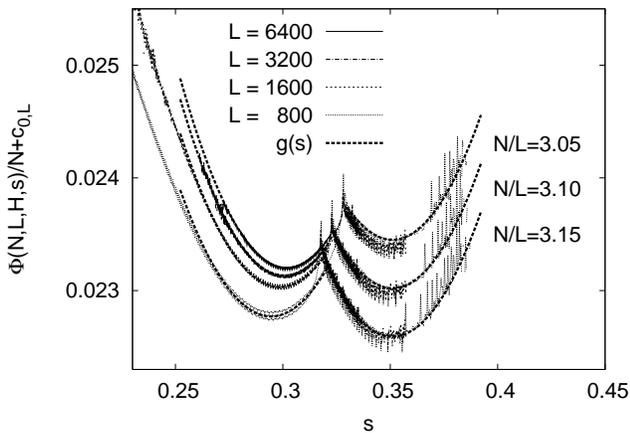, width=6.0cm, angle=270}
\caption{The Landau free energy divided by $N$, 
$\Phi(N,L,H,s)/N$, plotted
against $s$ for various values of $L$ and $H=28$. 
The two minima of $\Phi(N,L,H,s)/N$ are determined by fitting 
$g(s)=g_{\rm imp}(s)$, Eq.~(\ref{gimp}), in the imprisoned regime, and 
$g(s)=g_{\rm esc}(s)$, Eq.~(\ref{gesc}), in the escaped regime, 
going through those lower points around the two minima, respectively.} 
\label{fig-zs-fit}
\end{center}
\end{figure}

\begin{figure}
\begin{center}
\epsfig{file=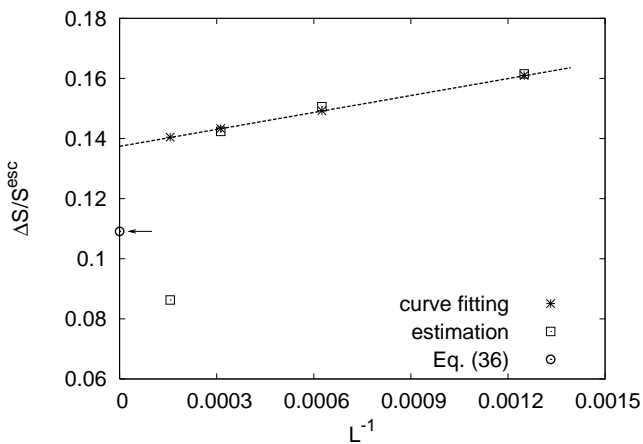, width=6.0cm, angle=270}
\caption{The reduced jump of the order parameter $\Delta S/S_{\rm esc}$
plotted against $L^{-1}$. }
\label{fig-f-ds}
\end{center}
\end{figure}

In order to determine the transition point and extract an accurate value for 
the jump in the order parameter from simulations, we use two parabolic functions 
$g_{\rm imp}(s)$ and $g_{\rm esc}(s)$ 
to fit the numerical data in the imprisoned and escaped regimes, respectively:
\be
   g_{\rm imp}(s)=a_{1,L}(s-s_{\rm eq,L}^{\rm imp})^2+ c_{1,L} \label{gimp}
\ee
and
\be
   g_{\rm esc}(s)=a_2(s-s_{\rm eq,L}^{\rm esc})^2+c_{2}+b_{2}\frac{N}{L}
\label{gesc}
\ee
where $a_{1,L}$, $c_{1,L}$, $a_2$, $c_{2}$, $b_2$, $s_{{\rm eq},L}^{\rm imp}$,
and $s_{\rm eq,L}^{\rm esc}$ are determined by curve fitting, and results
are shown in Table~\ref{table3} and Fig.~\ref{fig-zs-fit}.  
From the condition of equal depth of minima  
\be
g_{\rm imp}(s=s_{\rm eq,L}^{\rm imp})=g_{\rm esc}(s=s_{\rm eq,L}^{\rm esc}),
\ee
we obtain the transition points $(N/L)_{\rm tr}=3.13(2)$, $3.10(2)$, 
$3.09(1)$, and $3.08(1)$
for $L=800$, $1600$, $3200$, and $6400$, respectively, 
which are in perfect agreement with the results given by free energy,
Eq.~(\ref{f-nc-e}), and the results given by the variance of the end-to-end 
distance and the imprisoned monomers, Eq.~(\ref{f-nc-new-e}).

\begin{table}
\begin{center}
\caption{Results of the coefficients
$a_{1,L}$, $c_{1,L}$, $a_2$, $c_{2}$, $b_2$, $s_{{\rm eq},L}^{\rm imp}$,
and $s_{\rm eq,L}^{\rm esc}$ for the curve fitting in Fig.~\ref{fig-zs-fit}.}
\label{table3}
 \begin{ruledtabular}
 \begin{tabular}{rrrrrrrr}
 $L$    & $a_{1,L}$ & $ c_{1,L}$ & $s_{\rm eq,L}^{\rm imp}$ & $a_2$ & $c_{2}$ & $b_2$ 
& $s_{\rm eq,L}^{\rm esc}$  \\ \hline
   800 & 0.6281 & 0.02277 & 0.2945 &  0.6518 & 0.0497 & -0.0086 & 0.3510\\
  1600 & 0.6336 & 0.02303 & 0.2986 &         & & &\\
  3200 & 0.6675 & 0.02313 & 0.3007 &   & & & \\
  6400 & 0.6872 & 0.02320 & 0.3017 &   & & & \\
 \end{tabular}
 \end{ruledtabular}
 \end{center}
 \end{table}

The values for the reduced jump of the order parameter, 
\be
    \frac{\Delta S}{S_{\rm esc}}=
\frac{s_{\rm eq,L}^{\rm esc}-s_{\rm eq,L}^{\rm imp}}{s_{\rm eq,L}^{\rm esc}}
\ee
obtained by the curve fitting are plotted in Fig~\ref{fig-f-ds} against $L^{-1}$
together with the direct estimates in the simulations and the prediction by
the analytical theory, Eq.~(\ref{redS}).
We see that $\Delta S/S_{\rm esc}$ decreases as $L$ increase. 
As $L \rightarrow \infty$,
it remains finite and the value is slightly larger than the predicted value 
by the analytical theory. 
However, in view of the numerical uncertainties of our curve 
fitting we consider that the predictions of the analytical theory 
and the results by the MC simulations agree with each other quite well. 

\section{Summary and Discussion}
In this paper we attack the problem of the 2d-escape transition by combining 
several approaches. We first compare two simple pictures of the transition 
predicted for Gaussian chains and by a blob model. 
This comparison is useful from a 
general pedagogical point of view since the two models are in a sense 
complimentary: each captures some essential features of the phenomenon while 
failing in some other aspects. Both models are attractive because of their 
clarity, and although mathematically simple, lead to non-trivial results 
including finite-size effects in a phase transition. The third approach that 
was proposed in this paper attempts at incorporating the excluded volume 
effects in the framework of the Landau theory. We were not able to present 
an exact theory since it would require a detailed understanding of the 
end-to-end distribution of confined self-avoiding chains. To the best of 
our knowledge this problem is still not well explored. The simulations 
presented allowed us to evaluate the transition condition Eq.~(\ref{f-nc-e})
which represents the binodal line in the $(H, L/N)$ plane. It is of interest 
to extend the simulations in order to locate the spinodal lines where one of 
the states looses stability, and to construct the full phase diagram. It is 
also possible to explore the properties of metastable states and their 
life-times controlled by the barrier heights. It is clear from the results 
on the distribution of the order parameter, Fig.~\ref{fig-zs-org}, 
that the PERM algorithm 
experiences difficulties with sampling the configurations belonging 
to the escaped state, especially for long chains. The escaped branch of 
the distribution is cut-off quite sharply, which means that the important 
set of configurations characterized by larger stretching degree in the stem 
is vastly underrepresented. This is a generic problem that one encounters
when dealing with first-order transitions when the properties of the phases 
differ significantly. In our case, the PERM algorithm based on chain growth 
technique is perfectly tuned to generate homogeneous configurations of 
imprisoned chains but fails with strongly inhomogeneous escaped configurations. 
It is worth noting that a naive determination of the jumps in the average order 
parameter would have lead one to a wrong conclusion that the jump disappears 
in the thermodynamic limit. Again, we expect this to be a generic problem 
when simulating weak first-order transitions. The most reliable analysis of 
the nature of the transition would require a detailed examination of the order 
parameter distribution.

\section*{Acknowledgements}
We are grateful to the Deutsche Forschungsgemeinschaft (DFG)
for financial support: H.-P.H. was supported under grant NO
SFB 625/A3, while L.I.K. and A.M.S. received partial support 
under grants NO 436 RUS 113/863/0 and RFBR 05-03-32003-a. H.-P.H. thanks P. Grassberger and
W. Paul for very helpful discussions.

\section*{Appendix}

In this appendix, we discuss the finite-size behavior in the fluctuations
in the number of imprisoned monomers $N_{\rm imp}$
in more detail.
Following the technique of finite-size scaling analysis for
first-order transitions as described in~\cite{Binder84},
we write down the probability distribution of
the fraction of imprisoned monomers $m=N_{\rm imp}/N$
in the two-state model
\ba
    P(m)&=&\delta(m-m_1)\frac{e^{(t-t_{\rm tr})a}}{e^{(t-t_{\rm tr})a}+e^{-(t-t_{\rm tr})a}} 
\nonumber \\
&+&\frac{1}{\sqrt {2\pi } \sigma_0}e^{ - \frac{(m - m_2)^2 }{2\sigma_0^2 }} 
\frac{e^{-(t - t_{\rm tr} )a} }{e^{(t - t_{\rm tr} )a}  + e^{ - (t - t_{\rm tr} )a}}.
\ea
The first term accounts for the imprisoned state with $m$ strictly equal to 
$m_1=1$, while the second term describes the distribution of $m$ in the 
escaped state in the Gaussian approximation with the equilibrium average 
of $m$ equal to $m_2$ and dispersion $\sigma_0$; $t$ is the control 
parameter, $t_{\rm tr}$ 
is its critical value at the transition point, and $P(m)$ is normalized,
\be
    \int P(m)dm=1
\ee
At the transition point $t=t_{\rm tr}$,
\be
    P(m)=\frac{1}{2}\left[\delta(m-m_1)+\frac{1}{\sqrt {2\pi } 
\sigma_0}e^{ - \frac{(m - m_2)^2 }{2\sigma_0^2 }}\right]
\ee
which obeys the ``equal-weight rule'', while for $t \neq t_{\rm tr}$
the relative weight of the two states is $\exp\left[2(t-t_{\rm tr})a\right]$.
The constant $a^{-1}$ describes the range of $t$ over which the transition
is smeared out. For the Gaussian approximation to be meaningful
the dispersion of $m$ in the escaped state, $\sigma_0$, must be 
small compared to the difference $\Delta m = m_1-m_2$. Taking $t=L/N$ 
and using Eqs.~(\ref{Fimpl}), (\ref{Pesc}), and (\ref{Fescl}) of 
the Landau theory, one expects the following scaling: 
$a^{-1} \sim H^{2/3}/L$ 
and $\sigma_0^2 \sim H/L$.

Since the probability density is a sum of two contributions,
$P(m)=P_1(m)+P_2(m)$, the $k$-th moment of $m$ is defined by
\be
    <m^k>=\int m^k P(m) dm =<m^k>_1+<m^k>_2
\ee
where $<m^k>_{1,2}=\int m^k P_{1,2}(m) dm$.
Therefore, the first and second moment are given by
\be
   <m>=m_1 p_1 +m_2 p_2\;,
\ee
and
\be
   <m^2>=m_1^2 p_1 +(m_2^2+ \sigma_0^2)p_2\;,
\ee
here $p_1=e^{(t-t_{\rm tr})a}/{2\cosh\left[(t-t_{\rm tr})a\right]}$
is the relative weight of the imprisoned state, and $p_2=1-p_1$ is the
relative weight of the escaped state.

Instead of a $\delta$-function singularity at $t=t_{\rm tr}$,
the variance of the fraction of imprisoned monomers in a finite system becomes
\be
<m^2>-<m>^2=p_1 p_2 (\Delta m)^2 + p_2 \sigma_0^2 \label{sigma_tot}
\ee
which shows a smooth asymmetric peak close to $t=t_{\rm tr}$ of approximate height
$\Delta m^2+\sigma_0^2$. Here $\Delta m$ is
the relative reduction in the number of imprisoned monomers
at the transition point, for which the analytical Landau theory
predicts a value of $0.055$, see Eq.~(\ref{redN}).
The first term in Eq.~(\ref{sigma_tot}) is symmetric
with respect to the transition point since
$p_1 p_2 = 1/(4\cosh^2\left[(t-t_{\rm tr})a \right])$.
The second term, however, is asymmetric, as it describes the intrinsic
fluctuations in the escaped state.
The resultant asymmetry is clearly seen in Fig.~\ref{fig-d-m-0}.
We conclude that the full variance of $m$ is ill suited for a precise
determination of the transition point.

\begin{figure}
\begin{center}
\epsfig{file=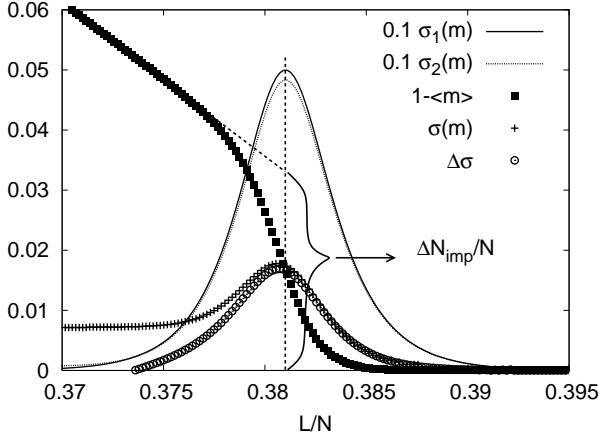, width=6.0cm, angle=270}
\caption{The square root of the variance
$\sigma_1(m)$ for the imprisoned states,
$\sigma_2(m)$ for the escaped chains,
$1-<m>$,
$\sigma(m)$ of the chain either in an imprisoned state
or in an escaped state, and the difference 
$\Delta \sigma = \sigma_1(m)-\sigma_2(m)$
against $L/N$.}
\label{fig-d-m-0}
\end{center}
\end{figure}

\begin{figure}
\begin{center}
\epsfig{file=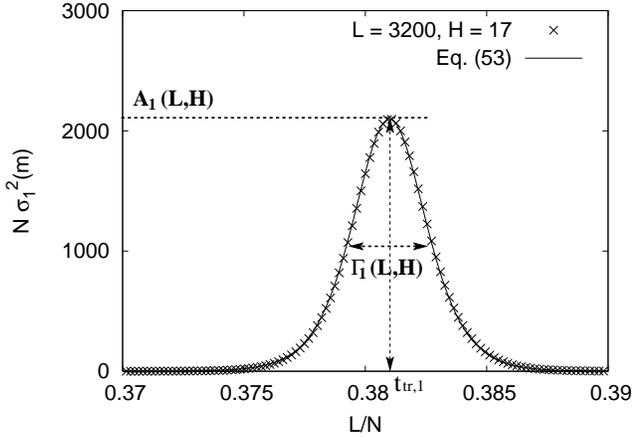, width=6.0cm, angle=270}
\caption{Variance due to the imprisoned configuration
multiplied by $N$, 
$N\sigma_1^2(m)$,
plotted against $L/N$ for $L=3200$ and $H=17$.
The solid curve is the best fit of Eq.~(\ref{sigma_1}), 
${Nm_1^2}/4{\cosh^2\left[(t-t_{\rm tr})a \right]}$,
with the height of the peak $A_{1}(L,H)=N(m_1)^2/4\approx2099.74$,
the FWHM $\Gamma_{\rm imp}(L,H)\approx 1.7627/a=0.0035$,
and the position of the peak $t_{\rm tr, 1}=(L/N)_{\rm tr, 1}=0.3810$.
The FWHM are given
by the distance between points on the curve shown
at which the corresponding height
reaches half height of the peaks (half maximum).
}
\label{fig-fit-m}
\end{center}
\end{figure}

\begin{figure}
\begin{center}
\epsfig{file=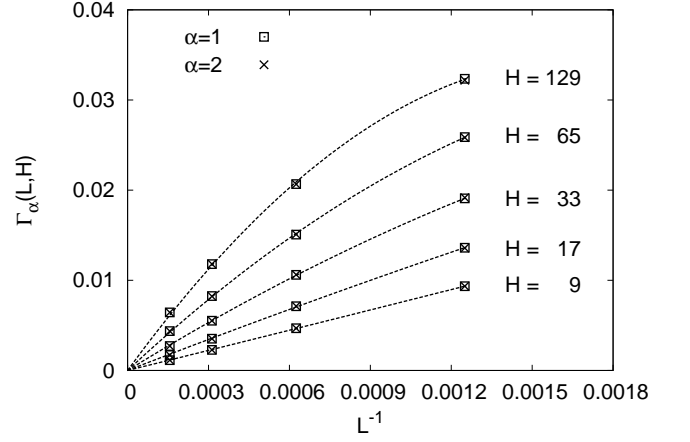, width=6.0cm, angle=270}
\caption{FWHM $\Gamma_\alpha(L,H)$ for the imprisoned state ($\alpha=1$)
and for the escaped state ($\alpha=2$) against $L^{-1}$.
The dashed curves are $a_{1,H}(H/L)+b_{1,H}(H/L)^2$ and give the best fit of the data.
Values of $a_{1,H}$ and $b_{1,H}$ are listed in Table~\ref{table4}.}
\label{fig-Gamma-m}
\end{center}
\end{figure}

The situation is quite different if we analyze the variances
calculated with the partial probability densities $P_1(m)$ and
$P_2(m)$ restricted to the imprisoned (escaped) configurations.
In the simulations, the product $N \sigma_{1,2}^2$ was calculated.
The variance due to the imprisoned configurations only (with $m_1=1$)
gives a perfectly symmetric curve as a function of the control
parameter:
\be
N\sigma_1^2(m) = Nm_1^2 p_1 p_2=\frac{N^2} 
{4\cosh^2\left[(t-t_{\rm tr})a \right]} \label{sigma_1}
\ee
with the peak value of $N_{\rm tr}/4 \approx LH^{1/3}/4$.
Numerical data presented in Fig.~\ref{fig-fit-m} supports
this prediction with very high accuracy.
The variance due to escaped configurations is somewhat modified by the 
intrinsic fluctuations in the escaped state
\ba
N\sigma_2^2(m) &=&N(m_2^2 p_1 p_2+ p_2 \sigma_0^2) \nonumber \\
 &\approx& Nm_2^2 p_1 p_2 \label{sigma_2}=\frac{N(1-\Delta m)^2}
{4\cosh^2\left[(t-t_{\rm tr})a\right]}
\ea
but in contrast to Eq.~(\ref{sigma_tot}), the asymmetric term is always
negligible. Indeed, the coefficient with the symmetric term,
$m_2^2=(1-\Delta m)^2$ is close to $1$ while both quantities 
$(\Delta m)^2$ and $\sigma_0^2$ are quite small.
In Fig.~\ref{fig-d-m-0}, we plot the full dispersion $\sigma(m)$
(the square root of the full variance) that includes contributions from
all configurations, and partial
dispersions $\sigma_1(m)$ and $\sigma_2(m)$ due to imprisoned
and escaped configurations separately, the difference
$\Delta \sigma= \sigma_1(m)-\sigma_2(m)$, as well as the average
fraction of escaped monomers, $1-<m>$, as functions of $L/N$.
It is clear that the full curve is strongly asymmetric in
contrast to partial dispersion curves, in good agreement with
the theoretical description above. 
On the other hand, the curve of $\Delta \sigma$ shows the same behaviour
as the curve of $\sigma(m)$ near the transition point, and the heights
of these two peaks correspond to the half size of the jump
$\Delta m=m_1-m_2=\Delta N_{\rm imp}/N$.
By fitting the partial 
variances $N\sigma_1^2(m)$ and $N\sigma_2^2(m)$ as functions
of $L/N$ according to Eqs.~(\ref{sigma_1}) and (\ref{sigma_2}),
respectively, we obtain the full width at half-maximum
(FWHM), 
$\Gamma(L,H)=2{\rm arccosh}(\sqrt{2})/a$, 
the height of the peak $A_\alpha(L,H)=Nm_\alpha^2/4$,
and the transition point $t_{\rm tr, \alpha}=(L/N)_{\rm tr, \alpha}$ for 
$\alpha=1$ (imprisoned configurations) and $\alpha=2$ (escaped configurations). 

\begin{table}
\begin{center}
\caption{Results of the coefficients $a_{1,H}$, $b_{1,H}$, $a_{2,H}$,
$c_{2,H}$ and $d_{2,H}$ for
the curve fitting in Fig.~\ref{fig-Gamma-m} and ~\ref{fig-A-m}.}
\label{table4}
 \begin{ruledtabular}
 \begin{tabular}{rrrrrr}
 $H$    & $a_{1,H}$ & $b_{1,H}$ & $a_{2,H}$ &  $c_{2,H}$ & $d_{2,H}$
\\ \hline
    9   & 0.8221 &  0.7841 & 0.9940 &  0.2131 & 2.2802 \\
   17   & 0.6845 & -2.0403 & 0.9845 &  0.0961 & 0.4540 \\
   33   & 0.5600 & -2.3531 & 0.9725 &  0.0430 & 0.0570 \\
   65   & 0.4301 & -1.3774 & 0.9545 &  0.0173 & 0.0095 \\
  129   & 0.3185 & -0.7332 & 0.9220 &  0.0070 & 0.0017
 \end{tabular}
 \end{ruledtabular}
 \end{center}
 \end{table}

\begin{figure*}
\begin{center}
$\begin{array}{c@{\hspace{0.1in}}c}
\multicolumn{1}{l}{\mbox{\large (a)}} &
        \multicolumn{1}{l}{\mbox{\large (b)}} \\ [-0.53cm]\\
\epsfig{file=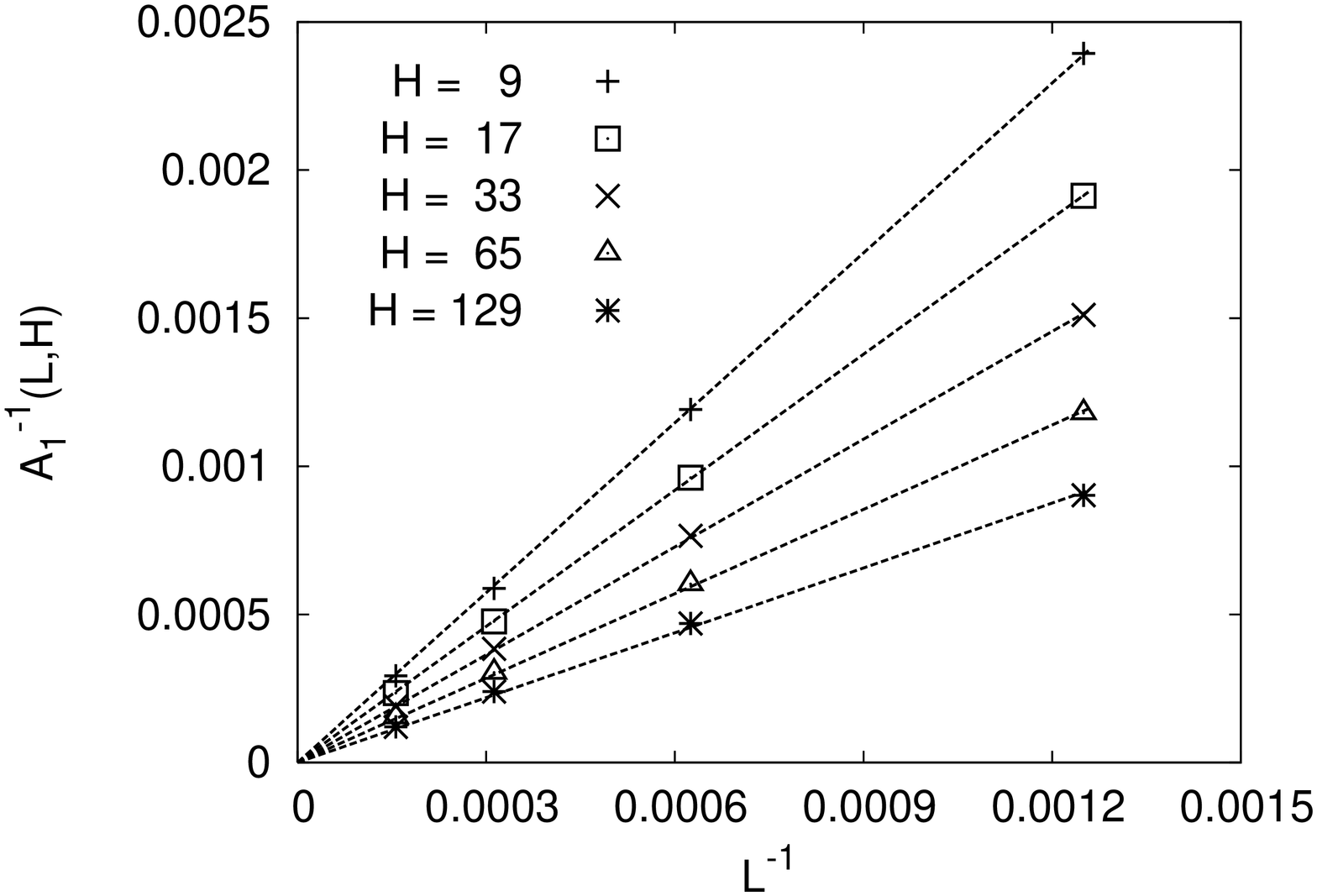, width=6.0cm, angle=270} &
\epsfig{file=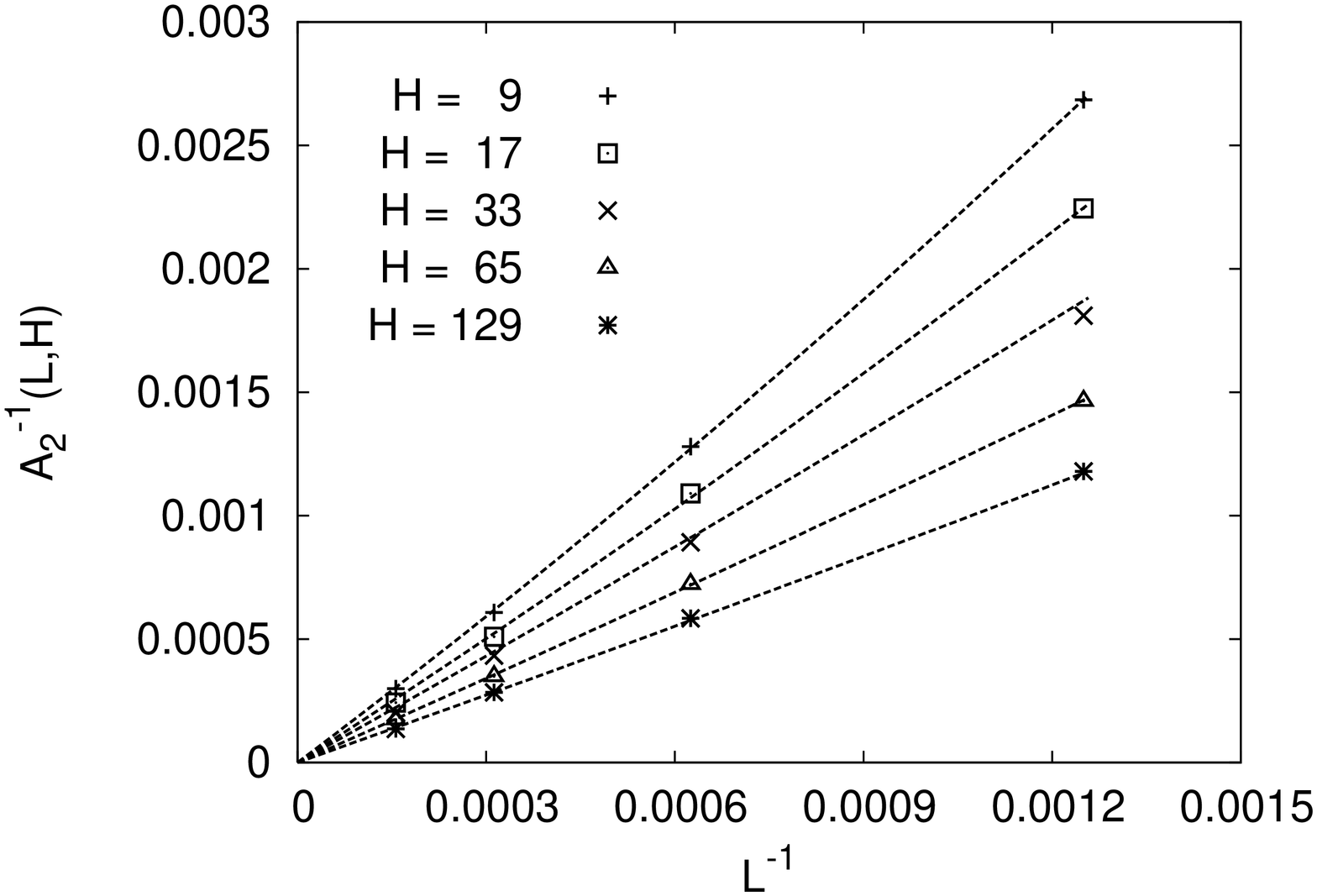, width=6.0cm, angle=270} \\
[0.4cm]
\end{array}$
\caption{Inverse of the height of the peaks for the imprisoned state
(a) $A_{1}^{-1}(L,H)$, and for the escaped state
(b) $A_{2}^{-1}(L,H)$, plotted
against $L^{-1}$. The dashed curves are (a) $a_{2,H}(4H^{-1/3}/L)$ and
(b) $c_{2,H}(H/L)+d_{2,H}(H/L)^2$, and give the best
fit of the data. Values of $a_{2,H}$, $c_{2,H}$, and $d_{2,H}$
are listed in Table~\ref{table4}.}
\label{fig-A-m}
\end{center}
\end{figure*}

\begin{figure}
\begin{center}
\epsfig{file=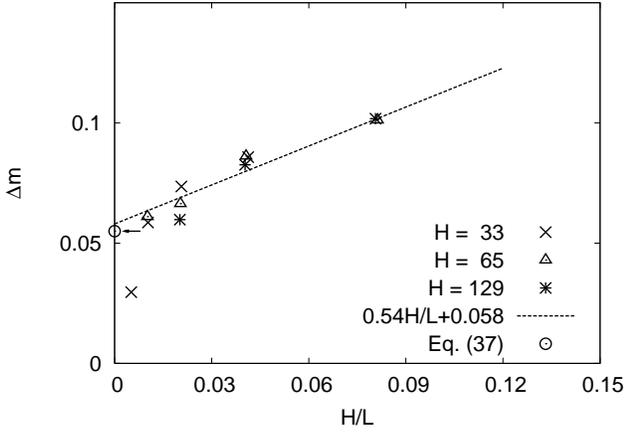, width=6.0cm, angle=270}
\caption{The relative reduction in the number of imprisoned monomers
$\Delta m$, plotted
against $H/L$.}
\label{fig-d-m1}
\end{center}
\end{figure}
 
One example of the curve fitting for $L=3200$ and $H=17$ is 
shown in Fig.~\ref{fig-fit-m}.
Note that the peak height and the transition point are related to
the theoretical prediction $A_1(L,H) \approx LH^{1/3}/4$ with
very high accuracy. 
Results of $\Gamma_\alpha(L,H)$, $A_\alpha(L,H)$
for $\alpha=1$
and for $\alpha=2$, and $t_{\rm tr,\alpha}$ are shown in 
Fig.~\ref{fig-Gamma-m}, ~\ref{fig-A-m} and \ref{fig-nc-new}.
In Fig.~\ref{fig-Gamma-m},
we see that the full widths $\Gamma_\alpha(L,H)$ for $\alpha=1$
and for $\alpha=2$ are overlapped with each other, and 
$\Gamma_\alpha \rightarrow 0$ as $1/L \rightarrow 0$ by fitting
the data using $a_{1,H}(H/L)+b_{1,H}(H/L)^2$. 
In Fig.~\ref{fig-A-m},  
the inverse of the height $A_1^{-1}(L,H) \rightarrow 0$ 
as $1/L\rightarrow 0$ by 
fitting the data using $a_{1,H}(4H^{-1/3}/L)$ and
$c_{1,H}(H/L)+d_{1,H}(H/L)^2$.
Since $\Gamma_\alpha \rightarrow 0$, and $A^{-1}_{\alpha} \rightarrow 0$
as $1/L \rightarrow 0$, i.e. a delta function, a sharp phase transition
occurs in the thermodynamic limit. It is a strong 
indication~\cite{Binder84} that
the transition is first-order like. Values of the coefficients 
$a_{1,H}$, $b_{1,H}$, $a_{2,H}$, $c_{2,H}$ and $d_{2,H}$
are listed in Table~\ref{table4}.

The relative reduction in the number of imprisoned monomers
\ba
\Delta m&=& m_1- m_2 \nonumber \\
&=& 2 (A_1^{1/2}-A_2^{1/2})/N^{1/2}
\ea
Results of $\Delta m$ for various values of $H$ and $L$
plotted against $H/L$ are shown in Fig.~\ref{fig-d-m1}. 
We see that there exist 
the systematic errors at small $H/L$.
Finally we obtain $\Delta m \approx 0.058$ 
at $H/L \rightarrow 0$ by a curve fitting,
which is slightly larger then the prediction,
Eq.~(\ref{redN}).

\end{document}